\documentclass[12pt,leqno]{article}

\title{Coarse-grained distributions and superstatistics}

\usepackage{amsthm,amsmath,amssymb,a4wide,psfig,epsf}
\def\mb#1{\setbox0=\hbox{$#1$}\kern-.025em\copy0\kern-\wd0
\kern-0.05em\copy0\kern-\wd0\kern-.025em\raise.0233em\box0}

\begin{document}

\author{Pierre-Henri Chavanis}
\maketitle
\begin{center}
Laboratoire de Physique Th\'eorique, Universit\'e Paul Sabatier,\\
118, route de Narbonne, 31062 Toulouse, France\\ E-mail: {\it
chavanis{@}irsamc.ups-tlse.fr }

\vspace{0.5cm}
\end{center}

\begin{abstract}

We show an interesting connection between non-standard (non
Boltzmannian) distribution functions arising in the theory of violent
relaxation for collisionless stellar systems (Lynden-Bell 1967) and
the notion of superstatistics recently introduced by Beck \& Cohen
(2003). The common link between these two theories is the emergence of
coarse-grained distributions arising out of fine-grained
distributions. The coarse-grained distribution functions are written
as a superposition of Boltzmann factors weighted by a non-universal
function. Even more general distributions can arise in case of
incomplete violent relaxation (non-ergodicity). They are stable
stationary solutions of the Vlasov equation. We also discuss analogies
and differences between the statistical equilibrium state of a
multi-components self-gravitating system and the metaequilibrium (or
quasi-equilibrium) states of a collisionless stellar system.  Finally,
we stress the important distinction between entropies, generalized
entropies, relative entropies and $H$-functions. We discuss
applications of these ideas in two-dimensional turbulence and for
other systems with long-range interactions.

\end{abstract}

\section{Introduction}
\label{intro}

Recently, several researchers have questioned the ``universality'' of
the Boltzmann distribution in physics. This problem goes back to
Einstein himself who did not accept Boltzmann's principle $S=k\ln W$
on a general scope because he argued that the statistics of a system
$(W)$ should follow from its dynamics and cannot have a universal
expression \cite{pais,cohen}. In 1988, Tsallis introduced a generalized form of
entropy in an attempt to describe complex systems
\cite{tsallis}. This was the starting point for several
generalizations of thermodynamics, statistical mechanics and kinetic
theories (see, e.g., \cite{collpap}).  A lot of experimental and
numerical studies (in an impressive number of domains of physics) has
then shown that complex systems exhibit non-standard distributions and
that, in many cases, they can be fitted by Tsallis $q$-distributions
\cite{bib}. However, there also exists physical systems (like those
that we shall consider here) that are described neither by Boltzmann
nor by Tsallis distributions.

An important question is to understand {\it why} non-standard
distributions and generalized entropies emerge in a system. We have
argued that non-standard distributions arise when microscopic
constraints are in action \cite{gfp}. They sometimes appear as
``hidden constraints'' inaccessible to the observer. For ``simple
systems'', the energetically accessible microstates are {\it
equiprobable} and a standard combinatorial analysis leads to the
Boltzmann entropy. Then, the equilibrium distribution (most probable
macrostate) maximizes the Boltzmann entropy at fixed macroscopic
constraints (mass, energy,...). For ``complex systems'', the {\it a
priori} accessible microstates are {\it not} equiprobable, some being
even forbidden, contrary to what is postulated in ordinary statistical
mechanics. The non-equiprobability of microstates can be due to
microscopic constraints (of various origin) that affect the dynamics.
In certain cases, the microscopic constraints can be dealt with by
using a generalized form of entropy. In principle, this entropy $S=\ln
W'$ should be obtained from a counting analysis by assuming that the
microstates which satisfy the macroscopic constraints {\it and the
microscopic constraints} are equiprobable. An example of microscopic
constraints is provided by the Pauli exclusion principle in quantum
mechanics which prevents two fermions with the same spin to occupy the
same site in phase space. Because of this constraint, the Boltzmann
entropy is replaced by the Fermi-Dirac entropy which puts a bound
$f({\bf x},{\bf v})\le\eta_{0}$ on the maximum value of the
distribution function.  In this example, the exclusion principle is
explained by quantum mechanics so it has a fundamental origin.
Another example is when the particles are subject to an excluded
volume constraint. In simplest models (e.g., a lattice model), this is
accounted for by introducing a Fermi-Dirac type entropy in physical
space which puts a bound $\rho({\bf x})\le
\sigma_{0}$ on the maximum value of the spatial density. 
These entropies can be obtained from a combinatorial analysis which
carefully takes into account the fact that two particles cannot be in
the same microcell in phase space or in physical space. More
generally, we can imagine other situations where some microscopic
constraints (not necessarily of fundamental origin) act on the system
and lead to non-standard forms of distribution functions and
entropies.

Non-Boltzmannian distributions can also emerge when the system does
not mix well (for some reason) so that the evolution is {\it
non-ergodic}. In that case, the system does not sample the a priori
energetically accessible phase space uniformly and prefers some
regions more than others. The effectively accessible phase space can
have a complicated geometrical structure. In many cases, we do not
know the nature of the microscopic constraints perturbing the
dynamics, so that they act as ``hidden constraints'' inaccessible to
the observer. We just see their effect indirectly because they lead to
non-standard distributions.  The fact that we do not know these
microscopic constraints implies an indetermination in the selection of
the entropy functional.  For example, the Tsallis entropies
\cite{tsallis} can be relevant for a certain type of non-ergodic behaviour 
when the phase space has a fractal or multifractal structure. This
is appropriate in particular for porous media and in the case of weak chaos. In
Tsallis generalized thermodynamics, the complexity of mixing is
encapsulated in a {\it single} parameter $q$ which indexes the
entropies and characterizes the degree of mixing ($q=1$ if the
evolution is ergodic). In some cases, it is possible to determine the
parameter $q$ directly from the microscopic dynamics. In more
complicated situations, it has to be adjusted to the situation by a
fit.  It would be interesting to obtain Tsallis form of entropy
directly from a counting analysis by assuming that the energetically
accessible microstates are equiprobable on a fractal phase space. In
that case, Tsallis entropy could be viewed as an entropy on a fractal.
One interesting aspect of Tsallis entropy is that it exhibits
mathematical properties very close to those possessed by the Boltzmann
entropy. Therefore, it represents the most natural extension of the
Boltzmann entropy to the case of ``complex'' systems.  However,
Tsallis entropy is not expected to describe all types of complex
systems. Depending on the constraints acting on the underlying
dynamics, there exists situations in which the observed distribution
differs from a $q$-distribution. In that case, we must consider more
general forms of entropy $S=-\int C(f)d{\bf x}d{\bf v}$ where $C(f)$
is a convex function \cite{gfp}. 

Several microscopic models have been constructed to show how
non-standard distributions and generalized entropies can emerge in a
system. By introducing a kinetic interaction principle (KIP),
Kaniadakis \cite{kaniadakis} has obtained a generalized form of
Boltzmann and Fokker-Planck equations that lead to a wide class of
distribution functions at equilibrium. These generalized equations
arise when the expression of the transition probabilities is more
general than usually considered. This can take into account quantum
statistics or non-ideal effects (e.g. excluded volume) that are
ignored in the standard derivation of the Boltzmann and Fokker-Planck
equations. On the other hand, Borland \cite{borland} and Chavanis
\cite{gfp} have introduced generalized stochastic processes and
generalized Fokker-Planck equations in which the diffusion coefficient
and the friction/drift terms explicitly depend on the concentration of
particles. The dynamics of particles described by these stochastic
processes has a complex (non-ergodic) phase space structure. These
equations lead to non-standard distributions at equilibrium and they
are associated with generalized free energy functionals which play the
role of Lyapunov functions. Generalized Fokker-Planck equations have
also been studied by Frank \cite{frank}. In fact, as discussed in
Chavanis \cite{gfp}, it is possible to generalize the usual kinetic equations
(Boltzmann, Landau, Kramers, Smoluchowski,...)  in such a way that
they satisfy a H-theorem for an arbitrary form of entropy. Boltzmann,
Fermi-Dirac, Bose-Einstein and Tsallis entropies are just special
cases of this general formalism. As indicated previously, the
generalization of standard kinetic models can be viewed as a heuristic
attempt to take into account ``hidden constraints'' in complex
systems. What we are doing, essentially, is to develop an effective
thermodynamical formalism (E.T.F.) to accommodate from our lack of
complete information on the microscopic dynamics of a complex system.

In a different context, Beck \& Cohen \cite{bc} have shown how
non-standard distributions can arise in a system if an external
variable (e.g. the temperature) is allowed to fluctuate. The
probability of energy $E$ is then given by a Laplace transform
$P(E)=\int_{0}^{+\infty}f(\beta)e^{-\beta E}d\beta$ where
$f(\beta)$ is the distribution of fluctuations that must be
regarded as given. When $f(\beta)$ is strongly peaked around a
temperature $\beta_{0}$, the Boltzmann distribution $P(E)={1\over
Z}e^{-\beta_{0} E}$ is recovered. Beck \& Cohen gave particular
examples of non-standard distributions $P(E)$ arising from this
formalism and Tsallis \& Souza \cite{souza} constructed the
generalized entropies associated with these non-standard
distributions.

At the same time (ignoring the works of Beck \& Cohen and Tsallis
\& Souza), we revived the concept of violent relaxation introduced
by Lynden-Bell \cite{lb} for collisionless stellar systems described
by the Vlasov-Poisson system and we showed how this theory predicts
metaequilibrium states characterized by non-standard distribution
functions
\cite{gfp,aa3}. Assuming complete relaxation (ergodicity), the
coarse-grained distribution function (DF) is given by
$\overline{f}(\epsilon)={1\over Z(\epsilon)}\int_{0}^{+\infty}
\chi(\eta)\eta e^{-\eta(\beta\epsilon+\alpha)}d\eta$ where the
function $\chi(\eta)$ accounts for the conservation of the Casimir
integrals and is determined by the initial conditions. In this
context, the Casimir integrals play the role of ``hidden
constraints'' because they are not accessible at the coarse-grained
scale (which is the scale of observation). Due to the Liouville
theorem in $\mu$-space, they can give rise to an effective ``exclusion
principle'' similar to the Pauli principle in quantum mechanics
\cite{lb,cs}. In particular, the coarse-grained distribution is
bounded by the maximum value of the initial (fine-grained)
distribution: $\overline{f}({\bf x},{\bf v},t)\le {\rm max}_{{\bf x},{\bf v}} 
\lbrace f({\bf x},{\bf v},t=0)\rbrace $.  We gave
particular examples of non-standard distributions
$\overline{f}(\epsilon)$ arising from this formalism, with emphasis on
the Fermi-Dirac distribution
\cite{cs}, and we introduced the notion of ``generalized
entropies'' $S[\overline{f}]=-\int C(\overline{f})d{\bf x}d{\bf v}$
(in $\overline{f}$-space) associated with these coarse-grained
distributions. The same ideas apply in two-dimensional (2D) turbulence
where the coarse-grained vorticity is given by
$\overline{\omega}(\psi)={1\over Z(\psi)}\int_{-\infty}^{+\infty}
\chi(\sigma)\sigma e^{-\sigma(\beta\psi+\alpha)}d\sigma$
\cite{miller,csr,houches}.  In the case of geophysical flows that are
forced at small scale, Ellis {\it et al.} \cite{ellis} interpret
$\chi(\sigma)$ as a prior vorticity distribution encoding the
statistics of forcing while for freely evolving flows $\chi(\sigma)$
is determined from the initial conditions by the Casimirs.  In the
point of view of Ellis {\it et al.} \cite{ellis}, further discussed in
Chavanis \cite{geo}, the function $\chi(\sigma)$ must be regarded as
{\it given} and it directly determines the form of generalized entropy
$S[\overline{\omega}]=-\int C(\overline{\omega})d{\bf x}$ (in
$\overline{\omega}$-space) associated with the coarse-grained
vorticity field. The small-scale forcing, encapsulated in the function
$\chi(\sigma)$, can be viewed as a ``hidden constraint'' which affects
the structure of the coarse-grained vorticity.

The object of this paper is to emphasize the similarity between
the Beck-Cohen superstatistics and the coarse-grained distributions
arising in theories of violent relaxation. The point of
superstatistics is that experimentally or numerically observed
distributions are in general coarse-grained distributions which arise
as averages of finer-grained distributions. Therefore, Lynden-Bell's
statistics is a sort of superstatistics. This connection has not been
noted previously and we think that it deserves to be pointed out in
detail. Furthermore, the notion of generalized entropies that we gave in
\cite{gfp} in the context of the theory of violent relaxation
is similar to that given by Tsallis \& Souza
\cite{souza} in relation with the Beck-Cohen superstatistics.  

The paper is organized as follows. We first start to emphasize the
distinction between the statistical equilibrium state of a $N$-stars
system described by the Hamilton equations and the metaequilibrium
states of a collisionless stellar system described by the Vlasov
equation. To stress the analogies and the differences, we consider a
stellar system with a distribution of mass. The statistical
equilibrium state is described in Sec. \ref{sec_stateq} and the theory
of violent relaxation is discussed in Sec. \ref{sec_meta}.  The
similarities (and differences) between coarse-grained distribution
functions and superstatistics is shown in Sec.
\ref{sec_super}. We introduce the notion of generalized entropy
$S[\overline{f}]$ associated with the coarse-grained distributions in Sec.
\ref{sec_ge}. We show that the generalized entropies associated
with the coarse-grained DF predicted by Lynden-Bell can {never} be the
Tsallis functional $S_{q}[\overline{f}]=-{1\over q-1}\int
(\overline{f}^{q}-\overline{f})d{\bf r}d{\bf v}$ because Lynden-Bell's
distribution is defined for all energies $\epsilon$ while Tsallis
$q$-distribution (with $q>1$) has a compact support (the distribution
function drops to zero at a finite energy). Then, in
Sec. \ref{sec_incomplete}, we insist on the notion of {\it incomplete}
violent relaxation and on the limitations of Lynden-Bell's statistical
prediction. As the fluctuations weaken as the system approaches
equilibrium, it can be trapped in a stationary solution of the Vlasov
equation which is not the most mixed state. We interpret Tsallis
functional $S_{q}[\overline{f}]$ as a particular $H$-function in the
sense of Tremaine, H\'enon \& Lynden-Bell \cite{thlb}, not as an
entropy. We show that the proper form of Tsallis entropy in the
context of violent relaxation is a functional $S_{q}[\rho]=-{1\over
q-1}\int (\rho^{q}-\rho)d\eta d{\bf r}d{\bf v}$ of the fine-grained
distribution $\rho({\bf r},{\bf v},\eta)$. The maximization of
$S_{q}[\rho]$ at fixed mass, energy and Casimirs is a condition of
thermodynamical stability (in a generalized sense). By contrast, the
maximization of a $H$-function (e.g., the Tsallis $H$-function) at
fixed mass and energy is a condition of nonlinear dynamical stability
for a steady state of the Vlasov-Poisson system of the form
$\overline{f}=\overline{f}(\epsilon)$ with $\overline{f}'(\epsilon)<0$
\cite{gfp,aa3,cst}.  The $H$-functions can be used to construct a wide
class of stable models of galaxies which can be an alternative to
Lynden-Bell's prediction in case of incomplete relaxation. Another
alternative is to develop a {\it dynamical} theory of violent
relaxation \cite{csr,mn} in order to understand what limits mixing. In
that case, non-ergodicity is explained as a decay of the fluctuations
of the gravitational field driving the relaxation, not by a complex
structure of phase space. Generalized entropies like $S_{q}[\rho]$ or
$C(\rho)$ are not necessary in that approach. Finally, in
Sec. \ref{sec_turb}, we discuss these ideas in the context of 2D
turbulence and show that the notions of prior vorticity distributions
and relative entropies introduced by Ellis {\it et al.}
\cite{ellis} make the analogies with superstatistics much closer than
for freely evolving systems.

\section{Statistical equilibrium state of a multi-components
stellar system} \label{sec_stateq}

We wish to determine the statistical equilibrium state of a
stellar system made of stars with different mass $m_{i}$. This
Hamiltonian system is described by the microcanonical ensemble
where the energy $E$ and the particle numbers $N_{i}$ (for each
species) are fixed. A thermal equilibrium state is established due
to the development of stellar encounters which randomize the
distribution of particles (``collisional'' mixing).
Mathematically, this statistical equilibrium state is obtained
when the infinite time limit $t\rightarrow +\infty$ is taken
before the thermodynamic limit $N\rightarrow +\infty$ 
defined in \cite{new,meta}. This statistical approach is adapted to the case
of globular clusters whose age is of the same order as the
Chandrasekhar relaxation time $t_{relax}\sim (N/\ln N) t_{D}$
\cite{bt}. We shall determine the most probable distribution of
stars at statistical equilibrium by using a combinatorial
analysis, assuming that all accessible microstates (with given $E$
and $M_{i}=N_{i}m_{i}$) are equiprobable. To that purpose, we
divide the $\mu$-space $\lbrace {\bf r},{\bf v}\rbrace$ into a
very large number of microcells with size $h$. We do not put any
exclusion, so that a microcell can be occupied by an arbitrary
number of particles.  We shall now group these microcells into
macrocells each of which contains many microcells but remains
nevertheless small compared to the phase-space extension of the
whole system. We call $\nu$ the number of microcells in a
macrocell. Consider the configuration $\lbrace n_{ij} \rbrace$
where $n_{ij}$ is the number of particles of species $j$ in the
macrocell $i$. Using the standard combinatorial procedure
introduced by Boltzmann, the probability of the state $\lbrace
n_{ij}\rbrace$, i.e. the number of microstates corresponding to
the macrostate $\lbrace n_{ij}\rbrace$, is given by
\begin{equation}
\label{bol1} W(\lbrace n_{ij}\rbrace)=\prod_{i,j}N_{j}!{\nu^{n_{ij}}\over n_{ij}!}.
\end{equation}
This is the Maxwell-Boltzmann statistics. As is customary, we define
the entropy of the state $\lbrace n_{ij} \rbrace$ by
\begin{equation}
S(\lbrace n_{ij} \rbrace)=\ln W(\lbrace n_{ij} \rbrace). \label{bol2}
\end{equation}
It is convenient here to return to a representation in terms of the
distribution function giving the phase-space density of species $j$ in
the $i$-th macrocell: $f_{ij}=f_{j}({\bf r}_i,{\bf
v}_i)={n_{ij}m_{j}/\nu h^{3}}$. Using the Stirling formula $\ln
n!=n\ln n-n$, we have
\begin{equation}
\ln W(\lbrace n_{ij} \rbrace)=-\sum_{i,j}n_{ij}\ln {n_{ij}}=-\sum_{i,j}\nu h^{3}{f_{ij}\over m_{j}}\ln {f_{ij}\over m_{j}}. \label{bol3}
\end{equation}
Passing to the continuum limit
$\nu\rightarrow 0$, we obtain the usual expression of the Boltzmann
entropy for different types of particles
\begin{equation}
\label{bol4} S_{B}=-\sum_{i}\int {f_{i}\over m_{i}}\ln {f_{i}\over m_{i}} d^{3}{\bf r}d^{3}{\bf v},
\end{equation}
up to some unimportant additive constant. This is the expression used
by Lynden-Bell \& Wood \cite{lbw} in their thermodynamical description
of ``collisional'' stellar systems (globular clusters). Assuming
ergodicity, the statistical equilibrium state, corresponding to the
most probable distribution of particles, is obtained by maximizing the
Boltzmann entropy (\ref{bol4}) while conserving the mass of each
species
\begin{equation}
\label{bol5} M_{i}=\int f_{i}d^{3}{\bf r}d^{3}{\bf v},
\end{equation}
and the total energy
\begin{equation}
\label{bol6} E={1\over 2}\int f v^{2}d^{3}{\bf r}d^{3}{\bf v}+{1\over 2}\int \rho\Phi d^{3}{\bf r}
\end{equation}
where $f({\bf r},{\bf v})=\sum_{i}f_{i}({\bf r},{\bf v})$ is the total distribution function and $\rho=\int f d^{3}{\bf v}$ the total density. The gravitational potential is determined by the Poisson equation
\begin{equation}
\label{bol7} \Delta\Phi=4\pi G\rho.
\end{equation}
Introducing Lagrange multipliers and writing the variational principle
in the form
\begin{equation}
\label{bol8} \delta S_{B}-\beta\delta E-\sum_{i}\alpha_{i}\delta M_{i}=0,
\end{equation}
we get
\begin{equation}
\label{bol9} f_{i}=A_{i}e^{-\beta m_{i}({v^{2}\over 2}+\Phi)}.
\end{equation}
The total distribution function is therefore given by
\begin{equation}
\label{bol10} f=\sum_{i} A_{i}e^{-\beta m_{i}({v^{2}\over 2}+\Phi)}.
\end{equation}
It is a superposition of Maxwell-Boltzmann distributions with
equal temperature $k_B T=1/\beta$ and different mass $m_{i}$.
According to the theorem of equipartition of energy, the mean
squared velocity of species $i$ decreases with mass such that
\begin{equation}
\label{bol11} \langle v^{2}\rangle_{i}={\int e^{-\beta
m_{i}{v^{2}\over 2}}v^{2}d^{3}{\bf v}\over \int e^{-\beta
m_{i}{v^{2}\over 2}}d^{3}{\bf v}}={3k_B T\over m_{i}}.
\end{equation}
Therefore, heavy particles have less velocity dispersion to resist
gravitational attraction so they preferentially orbit in the inner
region of the system. This leads to mass segregation. The effect
of mass segregation can also be appreciated by writing the
distribution function (\ref{bol9}) in the form
\begin{equation}
\label{bol12}
f_{i}(\epsilon)=C_{ij}\lbrack f_{j}(\epsilon)\rbrack^{m_{i}/m_{j}},
\end{equation}
where $C_{ij}=A_i/A_j^{m_i/m_j}$ is a constant independent on the
individual energy $\epsilon=v^{2}/2+\Phi$. On the other hand,
developing a kinetic theory for a multi-components
self-gravitating system, one obtains the multi-species Landau equation
\begin{equation}
\label{bol13} {\partial f_{i}\over\partial t}+{\bf v}\cdot
{\partial f_i\over\partial {\bf r}}+{\bf F}\cdot {\partial
f_i\over\partial {\bf v}}={\partial\over\partial
v^{\mu}}\sum_{j}\int  K^{\mu\nu}\biggl (m_{j}f'_{j}{\partial
f_{i}\over\partial v^{\nu}}-m_{i}f_{i}{\partial
f'_{j}\over\partial v^{'\nu}}\biggr )d^{3}{\bf v}',
\end{equation}
\begin{equation}
\label{bol14}
K^{\mu\nu}=2\pi G^{2}{1\over u}\ln\Lambda \biggl (\delta^{\mu\nu}-{u^{\mu}u^{\nu}\over u^{2}}\biggr ),
\end{equation}
where ${\bf u}={\bf v}-{\bf v}'$ is the relative velocity of the
particles involved in an encounter, $\ln
\Lambda=\int_{0}^{+\infty}dk/k$ is the Coulomb factor (regularized
with appropriate cut-offs) and we have set $f_{j}'=f_{j}({\bf r},{\bf v}',t)$
assuming that the collisions can be treated as local (see Kandrup
\cite{kandrup} for a critical discussion of this approximation and
formal generalizations). The Landau-Poisson system conserves the total
mass of each species of particles and the total energy of the
system. It also increases the Boltzmann entropy (\ref{bol4})
monotonically: $\dot S_{B}\ge 0$ (H-theorem).  The linearly
dynamically stable stationary solutions of the Landau-Poisson system
are determined by the mean-field Maxwell-Boltzmann distributions
(\ref{bol9}) which are local maxima of the Boltzmann entropy at fixed
$E$, $N_{i}$, so they correspond to statistical equilibrium states. We
emphasize that the Boltzmann distribution is the {\it only} stationary
solution of the Landau equation. The problems linked with the absence
of strict statistical equilibrium state in self-gravitating systems
and the notion of long-lived metastable states are discussed in
\cite{meta}.

\section{Violent relaxation of collisionless stellar systems}
\label{sec_vr}

\subsection{The Vlasov-Poisson system}
\label{sec_vp}

We shall now contrast the statistical equilibrium state of
``collisional'' stellar systems (globular clusters) to the
metaequilibrium, or quasi-equilibrium,  states of
``collisionless'' stellar systems (elliptical galaxies). The
distinction between collisional and collisionless dynamics is just
a question of timescales. The age of elliptical galaxies is by
many orders of magnitude smaller than the Chandrasekhar relaxation
time \cite{bt} so that their evolution  is governed by the
Vlasov-Poisson system
\begin{equation}
\label{vp1}
{\partial f\over\partial t}+{\bf v}\cdot {\partial f\over\partial {\bf r}}+{\bf F}\cdot {\partial f\over\partial {\bf v}}=0,
\end{equation}
\begin{equation}
\label{vp2}
\Delta\Phi=4\pi G\int f d^{3}{\bf v},
\end{equation}
where ${\bf F}=-\nabla\Phi$ is the force by unit of mass
experienced by a particle. Mathematically, the Vlasov equation is
obtained when the $N\rightarrow +\infty$ limit is taken before the
$t\rightarrow +\infty$ limit. Indeed, the collision term in Eq.
(\ref{bol13}) scales as $1/N$ in a proper thermodynamic limit
\cite{new} so that it vanishes for $N\rightarrow +\infty$. The
Vlasov equation, or collisionless Boltzmann equation, simply
states that, in the absence of encounters, the distribution
function $f$ is conserved by the flow in phase space. This can be
written $df/dt=0$ by using the advective derivative. The Vlasov
equation can also be obtained from the $N$-body Liouville equation
by making a mean-field approximation, i.e. the $N$-body
distribution factors out in a product of $N$ one-body
distributions. We note that the individual mass $m_{i}$ of the
stars does not appear in the Vlasov equation. Therefore, in the
collisionless regime, the evolution of the total distribution
function does not depend on how many species of particles exist in
the system (unlike the Landau equation).  This implies that the
collisionless dynamics does not lead to a segregation by mass
contrary to the collisional dynamics. It is easy to show that the
Vlasov equation conserves the total mass $M$ and the total energy
$E$ of the system. Furthermore, the Vlasov equation conserves an
infinite number of invariants called the Casimir integrals. They
are defined by $ I_{h}=\int h(f) d^{3}{\bf r} d^{3}{\bf v}$ for
any continuous function $h(f)$. The conservation of the Casimirs
is equivalent to the conservation of the moments of the
distribution function denoted
\begin{equation}
\label{vp3} M_n=\int f^{n} d^{3}{\bf r}d^{3}{\bf v}.
\end{equation}
The Vlasov-Poisson system also conserves angular momentum and impulse
but these constraints will not be considered here. Finally, the Vlasov
equation admits an infinite number of stationary solutions whose
general form is given by the Jeans theorem \cite{bt}.

\subsection{The metaequilibrium state}
\label{sec_meta}

The Vlasov-Poisson system develops very complex filaments as a
result of a mixing process in phase space (collisionless mixing).
In this sense, the fine-grained distribution function ${f}({\bf
r},{\bf v},t)$ will never reach a stationary state but will rather
produce intermingled filaments at smaller and smaller scales.
However, if we introduce a coarse-graining procedure, the
coarse-grained distribution function $\overline{f}({\bf r},{\bf
v},t)$ will reach a metaequilibrium state $\overline{f}({\bf
r},{\bf v})$ on a very short timescale, of the order of the
dynamical time $t_D$. This is because the evolution continues at 
scales smaller than the scale of observation (coarse-grained).
This process is known as ``phase mixing'' and ``violent
relaxation'' (or collisionless relaxation) \cite{bt}. Lynden-Bell
\cite{lb} has tried to predict the metaequilibrium state achieved
by the system in terms of statistical mechanics. This approach is of
course quite distinct from the statistical mechanics of the $N$-body
system (exposed in Sec. \ref{sec_stateq}) which describes the
statistical equilibrium state reached by a discrete $N$-body
Hamiltonian system for $t\rightarrow +\infty$. In Lynden-Bell's
approach, we make the statistical mechanics of a {\it field}, the
distribution function $f({\bf r},{\bf v},t)$ whose evolution is
governed by the Vlasov-Poisson system, while in Sec.  \ref{sec_stateq}
we made the statistical mechanics of a system of {\it point} particles
described by Hamilton equations. In the following, we shall summarize
the theory of Lynden-Bell and make the connection with the notion of
superstatistics.

Let $f_{0}({\bf r},{\bf v})$ denote the initial (fine-grained)
distribution function. We discretize $f_{0}({\bf r},{\bf v})$ in a
series of levels $\eta$ on which $f_{0}({\bf r},{\bf v})\simeq \eta$
is approximately constant. Thus, the levels $\lbrace \eta\rbrace$
represent all the values taken by the fine-grained distribution
function. If the initial condition is unstable, the distribution
function $f({\bf r},{\bf v},t)$ will be stirred in phase space (phase
mixing) but will conserve its values $\eta$ and the corresponding
hypervolumes $\gamma(\eta)=\int \delta(f({\bf r},{\bf
v},t)-\eta)d^{3}{\bf r}d^{3}{\bf v}$ as a property of the Vlasov
equation (this is equivalent to the conservation of the Casimirs). Let
us introduce the probability density $\rho({\bf r},{\bf v},\eta)$ of
finding the level of phase density $\eta$ in a small neighborhood of
the position ${\bf r},{\bf v}$ in phase space. This probability
density can be viewed as the local area proportion occupied by the
phase level $\eta$ and it must satisfy at each point the normalization
condition
\begin{equation}
\label{E1}
\int\rho({\bf  r,v},\eta)d\eta=1.
\end{equation}
The locally averaged (coarse-grained) distribution function is
then expressed in terms of the probability density as
\begin{equation}
\label{E2}
\overline{f}({\bf  r,v})=\int\rho({\bf  r,v}, \eta)\eta d \eta,
\end{equation}
and  the associated (macroscopic) gravitational potential satisfies
\begin{equation}
\label{E3}
\Delta{\overline{\Phi}}=4\pi G\int \overline{f} d^{3}{\bf v}.
\end{equation}
Since the gravitational potential is expressed by space integrals of
the density, it smoothes out the fluctuations of the distribution
function, supposed at very fine scale, so $\Phi$ has negligible
fluctuations (we thus drop the bar on $\Phi$). The conserved
quantities of the Vlasov equation can be decomposed in two groups. The
mass and energy will be called {\it robust integrals} because they are
conserved by the  coarse-grained distribution function:
$\overline{M[f]}=M[\overline{f}]$ and $\overline{E[f]}\simeq
E[\overline{f}]$. Hence
\begin{equation}
\label{E4}
M=\int \overline{f}d^{3}{\bf  r} d^{3}{\bf  v},
\end{equation}
\begin{equation}
\label{E5}
E=\int{1\over 2}\overline{f} v^{2} d^{3}{\bf  r}
 d^{3}{\bf  v}+{1\over 2}\int \overline{f} \ {\Phi} d^{3}{\bf  r} d^{3}{\bf  v}.
\end{equation}
As discussed above, the gravitational potential can be considered
as smooth, so we have expressed the energy in terms of the
coarse-grained fields $\overline{f}$ and ${\Phi}$ neglecting
the internal energy of the fluctuations $\overline{\tilde f\tilde
\Phi}$. Therefore, the mass and the energy can be calculated at any time of the evolution from the coarse-grained field $\overline{f}$. By contrast, the moments $M_{n}$ with $n\ge 2$ will be called
{\it fragile integrals} because they are altered on the coarse-grained
scale since $\overline{f^{n}}\neq \overline{f}^{n}$. Therefore, only
the moments of the fine-grained field
$M^{f.g.}_{n}=\overline{M_{n}[f]}=\int \overline{f^{n}}d^{3}{\bf r}d^{3}{\bf v}$ are conserved, i.e.
\begin{equation}
\label{E6}
M^{f.g.}_{n}=\int \rho({\bf  r,v},\eta)\eta^{n} d^{3}{\bf r}d^{3}{\bf v}d\eta.
\end{equation}
The moments of the coarse-grained field $M^{c.g.}_{n}[\overline{f}]=\int
\overline{f}^{n}d^{3}{\bf r}d^{3}{\bf v}$ are not conserved 
along the evolution since $M_{n}[\overline{f}]\neq
\overline{M_{n}[f]}$. In a sense, the moments $M^{f.g.}_{n}$ are
``hidden constraints'' because they are expressed in terms of the
fine-grained distribution $\rho({\bf r,v},\eta)$ and they cannot be
measured from the coarse-grained field. They can be only computed from
the initial conditions before the system has mixed or from the
fine-grained field. Since in many cases we do not know the initial
conditions nor the fine-grained field, they often appear as
``hidden''. Note that instead of conserving the fine-grained moments,
we can equivalently conserve the total hypervolume $\gamma(\eta)=\int
\rho d^{3}{\bf r}d^{3}{\bf v}$ of each level $\eta$.

After a complex evolution, we may expect the system to be in the most
probable, i.e. most mixed state, consistent with all the constraints
imposed by the dynamics (see, however, Sec. \ref{sec_incomplete}). We
define the mixing entropy as the logarithm of the number of
microscopic configurations associated with the same macroscopic state
characterized by the probability density $\rho({\bf r},{\bf
v},\eta)$. To get this number, we divide the macrocells $({\bf r},{\bf
r}+d{\bf r};{\bf v},{\bf v}+d{\bf v})$ into $\nu$ microcells of size
$h$ and denote by $n_{ij}$ the number of microcells occupied by the
level $\eta_{j}$ in the $i$-th macrocell. Note that a microcell can be
occupied only by one level $\eta_{j}$. This is due to the fact that we
make the statistical mechanics of a continuous field $f({\bf r},{\bf
v},t)$ instead of point mass stars as in Sec.
\ref{sec_stateq}. Therefore, we cannot ``compress'' that field,
unlike point-wise particles. A simple combinatorial analysis
indicates that the number of microstates associated with the
macrostate $\lbrace n_{ij}\rbrace$ is
\begin{equation}
W(\lbrace n_{ij}\rbrace)=\prod_{j}N_{j}!\prod_{i}{\nu!\over n_{ij}!},
\label{E7}
\end{equation}
where $N_{j}=\sum_{i}n_{ij}$ is the total number of microcells
occupied by $\eta_{j}$ (this is a conserved quantity equivalent to
$\gamma(\eta)$). We have to add the normalization condition
$\sum_{j}n_{ij}=\nu$, equivalent to Eq. (\ref{E1}) which prevents
overlapping of different levels (we note that we treat here the level
$\eta=0$ on the same footing as the others). This constraint plays
a role similar to the Pauli exclusion principle in quantum
mechanics. Morphologically, the Lynden-Bell statistics
(\ref{E7}) corresponds to a $4^{\rm th}$ type of statistics since
the particles are distinguishable but subject to an exclusion
principle \cite{lb}. There is no such exclusion for the
statistical equilibrium of point mass stars since they are free a
priori to approach each other, so we can put several particles in the 
same microcell.

Taking the logarithm of $W$ and using the Stirling formula, we
get
\begin{equation}
\ln W(\lbrace n_{ij} \rbrace)=-\sum_{i,j}n_{ij}\ln
{n_{ij}}=-\sum_{i,j}\nu h^{3}{\rho_{ij}}\ln {\rho_{ij}} \label{E8}
\end{equation}
where $\rho_{ij}=\rho({\bf r}_{i},{\bf v}_{i},\eta_{j})=n_{ij}/\nu h^{3}$.
Passing to the continuum limit
$\nu\rightarrow 0$, we obtain the Lynden-Bell mixing
entropy
\begin{equation}
S_{L.B.}[\rho]=-\int \rho({\bf r},{\bf v},\eta)\ln\rho({\bf r},{\bf v},\eta)d^{3}{\bf r}d^{3}{\bf v}d\eta.
\label{E9}
\end{equation}
Note that the Lynden-Bell entropy can be interpreted as the Boltzmann
entropy for a distribution of levels $\eta$ (including
$\eta=0$). Equation (\ref{E9}) is sometimes called a {\it
collisionless entropy} to emphasize the distinction with the {\it
collisional entropy} (\ref{bol4}) of Sec. \ref{sec_stateq}. Assuming
ergodicity or ``efficient mixing'' (which may not be realized in
practice, see Sec. \ref{sec_incomplete}), the statistical equilibrium
state is obtained by maximizing $S[\rho]$ while conserving mass $M$,
energy $E$ and all the Casimirs (or moments $M_n$). We need also to
account for the local normalization condition (\ref{E1}). This problem
is treated by introducing Lagrange multipliers, so that the first
variations satisfy
\begin{equation}
\delta S-\beta\delta E-\sum_{n\ge 1}\alpha_{n}\delta M_{n}-\int\zeta({\bf r},{\bf v})\delta\biggl (\int \rho({\bf r},{\bf v},\eta)d\eta\biggr )d^{3}{\bf r}d^{3}{\bf v}=0,
\label{E10}
\end{equation}
where ${ \beta}$ is the inverse temperature and ${\alpha }_{n}$ the
``chemical potential" associated with $M_n$. The resulting optimal
probability density is a Gibbs state which has the form
\begin{equation}
\label{E11} \rho({\bf r},{\bf v},\eta) ={1\over
Z}\chi(\eta)e^{-(\beta\epsilon+\alpha)\eta},
\end{equation}
where $\epsilon={v^2\over 2}+\Phi$ is the energy of a star by unit of
mass. In writing Eq. (\ref{E11}), we have distinguished the Lagrange
multipliers $\alpha$ and $\beta$ associated with the robust integrals
$M$ and $E$ from the Lagrange multipliers $\alpha_{n>1}$, associated
with the conservation of the fragile moments $M_{n>1}=\int \rho
\eta^{n}d\eta d^{3}{\bf r}d^{3}{\bf v}$, which have been regrouped in
the function $\chi(\eta)\equiv {\rm
exp}(-\sum_{n>1}\alpha_{n}\eta^{n})$. This distinction will make sense
in the following. Under this form, we see that the equilibrium
distribution of phase levels is a product of a universal Boltzmann
factor $e^{-(\beta\epsilon+\alpha)\eta}$ by a non-universal function
$\chi(\eta)$ depending on the initial conditions. The partition
function $Z$ is determined by the local normalization condition $\int
\rho d\eta=1$ leading to
\begin{equation}
\label{E12}
Z=\int \chi(\eta)e^{-\eta(\beta\epsilon+\alpha)}d\eta.
\end{equation}
Finally, the
equilibrium coarse-grained DF defined by $\overline{f}=\int
\rho \eta d\eta$ can be written
\begin{equation}
\label{E14}
\overline{f}={\int \chi(\eta)\eta e^{-\eta(\beta\epsilon+\alpha)}d\eta\over\int \chi(\eta) e^{-\eta(\beta\epsilon+\alpha)}d\eta},
\end{equation}
or, equivalently,
\begin{equation}
\label{E13}
\overline{f}=-{1\over\beta}{\partial\ln Z\over\partial \epsilon}=F(\beta\epsilon+\alpha)=\overline{f}(\epsilon).
\end{equation}
It is straightforward to check that this coarse-grained distribution
depending only on the energy $\epsilon$ is a stationary solution of
the Vlasov equation \cite{bt}.  Thus, for a given initial condition, the
statistical theory of Lynden-Bell selects a particular stationary
solution of the Vlasov equation (most mixed) among all possible ones
(an infinity!). Incidentally, the fact that the coarse-grained DF
should be a stationary solution of the Vlasov equation is not obvious;
this depends on the definition of coarse-graining, see \cite{cb}.
Specifically, the equilibrium state is obtained by solving the
differential equation
\begin{equation}
\label{diffeq}
\Delta\Phi=4\pi G\int f_{\alpha_{n},\beta}({v^{2}\over 2}+\Phi) d^{3}{\bf v},
\end{equation}
and relating the Lagrange multipliers $\alpha_{n}$, $\beta$ to the
constraints $M_{n}$, $E$. We note that the coarse-grained distribution
function $\overline{f}(\epsilon)$ can take a wide diversity of forms
depending on the function $\chi(\eta)$ determined by the fragile
moments (``hidden constraints''). Some examples will be given in
Sec. \ref{sec_super}. In the present context, the function
$\chi(\eta)$ is determined from the constraints {\it a
posteriori}. Indeed, we have to solve the full problem in order to get
the expression of $\chi(\eta)$. In this sense, the constraints
associated with the conservation of the fine-grained moments are
treated microcanonically. We emphasize that the function
$\overline{f}(\epsilon)$ depends on the detail of the initial
conditions unlike in ordinary statistical mechanics where only the
mass $M$ and the energy $E$ matter. Here, we need to know the value of
the fine-grained moments $M_{n}^{f.g.}$ which are accessible only in
the initial condition (or from the fine-grained field) since the {\it
observed} moments are altered for $t>0$ by the coarse-graining as the
system undergoes a mixing process ($M_{n}^{c.g.}\neq
M_{n}^{f.g.}$). This makes the practical prediction of
$\overline{f}(\epsilon)$ very complicated, or even impossible, since
we often do not know the initial conditions in detail (e.g., for the
formation of elliptical galaxies). In addition, in many cases, we
cannot be sure that the initial condition is not already mixed
(coarse-grained). If it has a fine-grained structure, this would
change a priori the prediction of the metaequilibrium state.

We note that the coarse-grained DF predicted by Lynden-Bell depends
only on the individual energy $\epsilon$ of the stars.  According to
the Jeans theorem \cite{bt}, such distribution functions form just a
particular class of stationary solutions of the Vlasov equation,
corresponding to spherical stellar systems (they even correspond to a
sub-class of spherical systems whose general distribution function
depends on energy $\epsilon$ and angular momentum ${\bf r}\times {\bf
v}$). From this simple fact, it is clear that the statistical theory
of violent relaxation is not able to account for the triaxial
structure of elliptical galaxies. More general stationary solutions of
the Vlasov equation can arise in case of incomplete violent relaxation
and they differ from Lynden-Bell's prediction (see
Sec. \ref{sec_incomplete}). We also note that $\overline{f}(\epsilon)$
is a monotonically decreasing function of energy. Indeed, from Eqs. (\ref{E11}) and
(\ref{E13}), it is easy to establish that
\begin{equation}
\label{g1} \overline{f}'(\epsilon)=-\beta f_{2},\qquad
f_{2}\equiv \int \rho (\eta-\overline{f})^{2}d\eta>0,
\end{equation}
where $f_{2}$ is the centered local variance of the distribution
$\rho({\bf r},{\bf v},\eta)$. Therefore, $\overline{f}'(\epsilon)\le
0$ since $\beta\ge 0$ is required to make the velocity profile
normalizable. Finally, the coarse-grained distribution function
satisfies $\overline{f}({\bf r},{\bf v})\le f_{0}^{max}$ where
$f_{0}^{max}$ is the maximum value of the initial (fine-grained)
distribution function. This inequality can be obtained from
Eq. (\ref{E14}) by taking the limit $\epsilon\rightarrow -\infty$ for
which $\overline{f}(\epsilon)\rightarrow \eta_{max}=f_{0}^{max}$ and
using the fact that $f(\epsilon)$ is a decreasing function. Of course,
the inequality $0\le \overline{f}\le f_{0}^{max}$ is clear from physical
considerations since the coarse-grained distribution function locally
averages over the fine-grained levels. Since the fine-grained
distribution function is conserved by the Vlasov equation, the
coarse-grained distribution function is always intermediate between the
minimum and the maximum values of $f_{0}$. Finally, we note that Lynden-Bell's
distribution (\ref{E14}) does not lead to a segregation by mass since
the individual mass of the particles does not appear in the Vlasov
equation on which the whole theory is based; however, it leads to a
segregation by phase levels $\eta$.

If the initial DF takes only two values $f_0=0$ and
$f_0=\eta_{0}$, the statistical prediction of Lynden-Bell for the
metaequilibrium state is
\begin{equation}
\label{E15}
\overline{f}={\eta_{0}\over 1+ e^{\eta_{0}(\beta\epsilon+\alpha)}},
\end{equation}
which is similar to the Fermi-Dirac distribution \cite{lb,cs}.  This
has to be contrasted from the statistical equilibrium state (for
$t\rightarrow +\infty$) of the one component self-gravitating gas
which is the Maxwell-Boltzmann distribution
\begin{equation}
\label{E16}
{f}=A e^{-\beta m\epsilon}.
\end{equation}
In the dilute limit of Lynden-Bell's theory $\overline{f}\ll \eta_{0}$
(which may be a good approximation for elliptical galaxies, see
\cite{lb}), the DF (\ref{E15}) becomes
\begin{equation}
\label{E17}
\overline{f}=A' e^{-\beta\eta_{0}\epsilon}.
\end{equation}
This is similar to the statistical equilibrium state (\ref{E16})
of the $N$-stars system. Therefore, in this approximation,
collisional and collisionless relaxation lead to similar
distribution functions (the Maxwell-Boltzmann distribution) but
with a completely different interpretation, corresponding to very
different timescales. To emphasize the difference, note in particular 
the bar on $f$ in Eq. (\ref{E17})
and the fact that the mass of the individual stars $m$ in (\ref{E16}) is
replaced by the value $\eta_{0}$ of the fine-grained distribution
function.

\subsection{Generalized entropies}
\label{sec_ge}

We have seen that the most probable local distribution of phase levels
$\rho({\bf r},{\bf v},\eta)$ maximizes the mixing entropy (\ref{E9})
while conserving mass, energy and all the fine-grained moments. This
functional of $\rho$ is the proper form of Boltzmann entropy in the
context of violent relaxation. It is obtained by a combinatorial
analysis taking into account the specificities of the collisionless
evolution. We shall now show that the most probable coarse-grained
distribution function $\overline{f}({\bf r},{\bf v})$ (which is the
function directly accessible to the observations) maximizes a certain
functional $S[\overline{f}]$ at fixed mass $M$ and energy $E$. This
functional of $\overline{f}$ will be called a ``generalized entropy''
(in a sense different to that given by Tsallis). It is non-universal
and depends on the initial conditions. It is determined indirectly by
the statistical theory of Lynden-Bell and cannot be obtained from a
combinatorial analysis, unlike $S[\rho]$. Such generalized
(non-Boltzmannian) functionals arise because they encapsulate the
influence of fine-grained constraints (Casimirs) that are not
accessible on the coarse-grained scale. They play the role of ``hidden
constraints'' in our general interpretation of non-standard entropies.
We note that the entropic functionals $S[\rho]$ and $S[\overline{f}]$
are defined on two different spaces. {\it The $\rho$-space is the relevant
one to make the statistical mechanics of violent relaxation
\cite{lb,csr}. The $\overline{f}$-space is a sort of 
projection of the $\rho$-space in the space of directly observable
(coarse-grained) distributions.}

Since the coarse-grained distribution function
$\overline{f}(\epsilon)$ predicted by the statistical theory of
Lynden-Bell depends only on the individual energy and is monotonically
decreasing, it extremizes a functional of the form
\begin{equation}
\label{g2} S[\overline{f}]=-\int C(\overline{f}) d^{3}{\bf r}d^{3}{\bf v},
\end{equation}
at fixed mass $M$ and energy $E$, where $C(\overline{f})$ is a convex function,
i.e. $C''>0$. Indeed, introducing Lagrange multipliers and writing the
variational principle as
\begin{equation}
\label{g3} \delta S-\beta \delta E-\alpha\delta M=0,
\end{equation}
we find that
\begin{equation}
\label{g4} C'(\overline{f})=-\beta\epsilon-\alpha.
\end{equation}
Since $C'$ is a monotonically increasing function of $\overline{f}$, we can
inverse this relation to obtain
\begin{equation}
\label{g5} \overline{f}=F(\beta\epsilon+\alpha)=\overline{f}(\epsilon),
\end{equation}
where
\begin{equation}
\label{g6} F(x)=(C')^{-1}(-x).
\end{equation}
From the identity
\begin{equation}
\label{g7} \overline{f}'(\epsilon)=-\beta/C''(\overline{f}),
\end{equation}
resulting from Eq. (\ref{g4}), $\overline{f}(\epsilon)$ is a
monotonically decreasing function of energy (if $\beta>0$). Thus,
Eq. (\ref{E13}) is compatible with Eq. (\ref{g5}) provided that we use 
the identification (\ref{g6}). Therefore, for any
function $F(x)$ determined by the function $\chi(\eta)$ in the
statistical theory, we can associate to the metaequilibrium state
(\ref{E13}) a generalized entropy (\ref{g2}) where $C(\overline{f})$
is given by Eq. (\ref{g6}) or equivalently by
\begin{equation}
\label{g8} C(\overline{f})=-\int^{\overline{f}}F^{-1}(x)dx.
\end{equation}
It can be shown furthermore that the coarse-grained distribution
(\ref{E13}) {\it maximizes} this generalized entropy at fixed energy
$E$ and mass $M$ (robust constraints) \footnote{This implies that
$\overline{f}$ is dynamically stable (nonlinearly) via the
Vlasov-Poisson system. Our discussion implicitly assumes 
that the system is confined within a box so as to avoid the infinite
mass problem (Sec. \ref{sec_incomplete}).}. We note that
$C(\overline{f})$ is a {\it non-universal} function which depends on
the initial conditions. Indeed, it is determined by the function
$\chi(\eta)$ which depends indirectly on the initial conditions
through the complicated procedures discussed in Sec.
\ref{sec_meta}. In general, $S[\overline{f}]$ is {\it not} the Boltzmann
functional $S_B[\overline{f}]=-\int \overline{f}\ln \overline{f}
d^{3}{\bf r}d^{3}{\bf v}$ (except in the dilute limit of the theory)
due to fine-grained constraints (Casimirs) that modify the form of
entropy that we would naively expect. This is why the metaequilibrium
state is described by non-standard distributions (even for an assumed
ergodic evolution).  The existence of ``hidden constraints'' (here the
Casimir invariants that are not accessible on the coarse-grained
scale) is the physical reason for the occurrence of non-standard
distributions and ``generalized entropies'' in our problem. {\it In
fact, the distribution is standard (Boltzmann-Gibbs) at the level of
the local distribution of fluctuations $\rho({\bf r},{\bf v},\eta)$
($\rho$-space) and non-standard at the level of the macroscopic
coarse-grained field $\overline{f}({\bf r},{\bf v})$ ($\overline{f}$-space).} 
We emphasize that the generalized entropies, which are maximized by
the coarse-grained distributions, are phenomenological in nature.  The
point here is that generalized entropies arise because we want to
phenomenologically extend the maximum entropy principle at the level
of coarse-grained distributions.

\subsection{Connection with superstatistics}
\label{sec_super}

We would like now to point out some connections between
coarse-grained distribution functions and superstatistics. Setting
$E\equiv \beta\epsilon+\alpha$, we can rewrite the ``partition
function'' (\ref{E12}) in the form
\begin{equation}
\label{s1}
Z(E)=\int_{0}^{+\infty} \chi(\eta)e^{-\eta E}d\eta.
\end{equation}
This is the Laplace transform of $\chi(\eta)$. Therefore, the
partition function $Z(E)=\hat\chi(E)$ can be used as a generating
function for constructing the moments of the fine-grained
distribution \cite{bellac,cp}. The coarse-grained distribution is
given by
\begin{equation}
\label{s2bis} \overline{f}(E)={1\over Z(E)}\int_{0}^{+\infty}
\chi(\eta)\eta e^{-\eta E}d\eta.
\end{equation}
We note that the Lynden-Bell statistics has a form similar to the
superstatistics $P(E)=\int_{0}^{+\infty}f(\beta)e^{-\beta E}d\beta$ of
Beck \& Cohen \cite{bc} provided that we identify the distribution of
temperature $f(\beta)$ to the distribution of phase levels
$\chi(\eta)$. Formally, the distribution $P(E)$ is expressed as a
Laplace transform like the partition function $Z(E)$. However,
physically, one should focus on the coarse-grained distribution
(\ref{s2bis}) as being {\it the} superstatistics in the present
context rather than the partition function (\ref{s1}). These
coarse-grained distributions do not exactly have the form considered
by Beck and Cohen, but this is a minor point. Superstatistics is an
idea foremost, not a proposition for a fixed form of average
distribution. The real point is that the coarse-grained distributions
do arise as averages (of some sort) of fine-grained distributions of
Boltzmann's type and so are superstatistics.

Due to these formal and physical analogies, we can transpose the
results of Beck \& Cohen \cite{bc} to the context of violent relaxation.
However, in the present case, the physical distribution is given
by
\begin{equation}
\label{s2}
\overline{f}(E)=-{\partial \ln Z\over\partial E},
\end{equation}
instead of $P(E)$. Therefore, for the same $f(\beta)$ and
$\chi(\eta)$, the distributions  $P(E)$ and $\overline{f}(E)$  will
differ because of this logarithmic derivative. In addition, we
must require that the distribution $\overline{f}(E)$ is
integrable, i.e. the spatial density $\rho=\int f d{\bf v}$ must
exist. We note finally that the generalized entropy associated to
the coarse-grained distribution $\overline{f}(E)$ is determined by
the relation
\begin{equation}
\label{s3}
C'(\overline{f})=-E,
\end{equation}
where the function $\overline{f}=\overline{f}(E)$ is specified by
Eq. (\ref{s2}) depending on $\chi(\eta)$. Therefore, $C(\overline{f})$
is obtained by inverting the relation $\overline{f}=-(\ln Z)'(E)$ and
integrating the resulting expression. In mathematical terms, we get
the nice formula defining the generalized entropy
\begin{equation}
\label{s3b} C(\overline{f})=-\int^{\overline{f}} \lbrack (\ln
\hat\chi)'\rbrack^{-1}(-x)dx.
\end{equation}
Interestingly, the notion of generalized entropy that we gave in the
context of violent relaxation in \cite{gfp} is similar to the one
given independently by Tsallis \& Souza \cite{souza} in the context of
superstatistics and by Almeida in the context of generalized
thermodynamics \cite{almeida}. Let us now consider particular examples
similar to those given by Beck \& Cohen \cite{bc}.  These examples are
given essentially to illustrate the fact that different forms of
non-standard distributions can emerge on the coarse-grained scale. We
do not claim that they have any particular physical
meaning (except (ii)). Furthermore, many other examples of
distributions and generalized entropies could be constructed.

{\it (i)  Uniform distribution:} We take $\chi(\eta)=1/b$ for $0\le\eta\le b$ and $\chi=0$ otherwise. Then
\begin{equation}
\label{s4}
Z(E)={1\over bE}(1-e^{-bE}),
\end{equation}
and
\begin{equation}
\label{s5}
\overline{f}(E)={1\over E}+{b\over 1-e^{bE}}.
\end{equation}
This distribution satisfies $\overline{f}(E)\rightarrow b$ for
$E\rightarrow -\infty$, $\overline{f}(0)=b/2$ and $\overline{f}(E)\sim
E^{-1}$ for $E\rightarrow +\infty$. Since $\overline{f}\sim v^{-2}$
for $v\rightarrow +\infty$, the density $\rho=\int f d{\bf v}$ exists
only in $d=1$ dimension.

{\it (ii) $2$-levels distribution:} We take $\chi(\eta)={1\over
2}\delta(\eta)+{1\over 2}\delta(\eta-b)$. Then
\begin{equation}
\label{s6}
Z(E)={1\over 2}(1+e^{-bE}),
\end{equation}
and
\begin{equation}
\label{s7}
\overline{f}(E)={b\over 1+e^{bE}}.
\end{equation}
This is similar to the Fermi-Dirac distribution \cite{lb,cs}. We
have $\overline{f}(E)\rightarrow b$ for $E\rightarrow -\infty$,
$\overline{f}(0)=b/2$ and $\overline{f}(E)\sim e^{-bE}$ for
$E\rightarrow +\infty$. Since $\overline{f}\sim e^{-b{v^{2}\over
2}}$ for $v\rightarrow +\infty$, the density $\rho=\int f d{\bf
v}$ exists in any dimension. Inverting the relation (\ref{s7}), we
get
\begin{equation}
\label{s8}
-E={1\over b}\biggl\lbrack \ln \overline{f}-\ln (b-\overline{f})\biggr \rbrack=C'(\overline{f}).
\end{equation}
After integration, we obtain
\begin{equation}
\label{s9} S[\overline{f}]=-\int \biggl\lbrace {\overline{f}\over
b}\ln {\overline{f}\over b}+\biggl (1-{\overline{f}\over b}\biggr
)\ln \biggl (1-{\overline{f}\over b}\biggr )\biggr\rbrace d^3{\bf
r}d^3{\bf v},
\end{equation}
which is similar to the Fermi-Dirac entropy. Note that for this
two-levels distribution, the generalized entropy (\ref{s9}) in
$\overline{f}$-space coincides with the mixing entropy (\ref{E9}) in
$\rho$-space since $\rho({\bf r},{\bf v},\eta)=(\overline{f}/
b)\delta(\eta-b)+(1-\overline{f}/ b)\delta(\eta)$. This is because the
distribution of phase levels $\rho({\bf r},{\bf v},\eta)=p({\bf
r},{\bf v})\delta(\eta-b)+p'({\bf r},{\bf v})\delta(\eta)$ can be
expressed in terms of the coarse-grained distribution function
$\overline{f}({\bf r},{\bf v})=p({\bf r},{\bf v})b$, using the
normalization condition $p+p'=1$. This is the only case where we have
the equivalence between the mixing entropy $S[\rho]$ and the
generalized entropy $S[\overline{f}]$. The fact that the `averaged'
Shannon entropy (\ref{E9}) and the generalized entropy (\ref{g2}) are
different in general has also been noted by Beck \cite{beckn} in a
different context.

{\it (iii) Gamma distribution:} We take
\begin{equation}
\label{s10}
\chi(\eta)={1\over b\Gamma(c)}\biggl ({\eta\over b}\biggr )^{c-1}e^{-\eta/b}
\end{equation}
with $c>0$ and $b>0$. Note that the case $c=1$ corresponds to the exponential distribution while $b\rightarrow +\infty$ corresponds to a power law. Then
\begin{equation}
\label{s11}
Z(E)=(1+{bE})^{-c}.
\end{equation}
As noted by Beck \& Cohen \cite{bc}, this is similar to Tsallis $q$-distribution (with $q<1$). However, in our context, the physical distribution is
\begin{equation}
\label{s12}
\overline{f}(E)={cb\over 1+bE}.
\end{equation}
It is defined only for $E>-1/b$. Furthermore, $\overline{f}(E)\sim
c E^{-1}$ for $E\rightarrow +\infty$ so that the spatial density
exists only in $d=1$ dimension. Inverting the relation
(\ref{s12}), we get
\begin{equation}
\label{s13}
-E={1\over b}\biggl (1-{cb\over  \overline{f}}\biggr ).
\end{equation}
After integration, we obtain
\begin{equation}
\label{s14}
C(\overline{f})={\overline{f}\over  b}-c\ln\overline{f}.
\end{equation}
Note that the first term can be absorbed in the Lagrange multiplier
$\alpha$ associated with the mass conservation so that the relevant
generalized entropy is
\begin{equation}
\label{s15} S\lbrack \overline{f}\rbrack=\int
\ln\overline{f}d^3{\bf r}d^3{\bf v}.
\end{equation}
It could be called the log-entropy. Note that when $E=v^{2}/2$, the
corresponding distribution function (\ref{s12}) is the Lorentzian. In a sense, the log-entropy can be viewed as a continuation of Tsallis entropy for $q=0$ (see Eq. (\ref{q})). This suggests to introducing the modified functional
\begin{equation}
\label{s15q} S_{q}\lbrack {f}\rbrack=-{1\over q-1}\int \biggl ({1\over q}f^{q}-f \biggr)d^{3}{\bf r}d^{3}{\bf v},
\end{equation}
which has properties similar to the Tsallis functional for $q\neq 0$
and which reduces to Eq. (\ref{s15}), leading to the distribution
(\ref{s12}), for $q\rightarrow 0$. More precisely, $S_{q}[f]=\int \ln
fd^{3}{\bf r}d^{3}{\bf v}+{K\over q}+O(q)$ for $q\rightarrow 0$, where
$K$ is a constant. Taking the variations $\delta
S_{q}$ at fixed mass and energy leads to
$f=(1-(q-1)E)^{1/(q-1)}$ which passes to the limit for $q\rightarrow
0$, unlike Eq. (\ref{q}).

{\it (iv) Gaussian distribution:} We take
\begin{equation}
\label{s15b}
\chi(\eta)=2\biggl ({\gamma\over\pi}\biggr) ^{1/2}e^{-\gamma \eta^{2}}
\end{equation}
with $\gamma>0$.  Then,
\begin{equation}
\label{s16}
Z(E)=e^{E^{2}\over 4\gamma}{\rm erfc}\biggl ({E\over 2\sqrt{\gamma}}\biggr ).
\end{equation}
The corresponding coarse-grained distribution can be written
\begin{equation}
\label{s17}
\overline{f}(E)={1\over \sqrt{\gamma}}H\biggl ({E\over 2\sqrt{\gamma}}\biggr ),\quad H(x)=x\biggl\lbrace {1\over \sqrt{\pi}xe^{x^{2}}{\rm erfc}(x)}-1\biggr\rbrace.
\end{equation}
This distribution satisfies $\overline{f}(E)\sim
-{E\over 2\gamma}$ for $E\rightarrow -\infty$ and  $\overline{f}(E)\sim
E^{-1}$ for $E\rightarrow +\infty$. The density exists only in
$d=1$ dimension.

In the examples considered above, only the Fermi-Dirac distribution
function is relevant for self-gravitating systems since the density
$\rho=\int f d{\bf v}$ is not defined for the others in $d=3$
dimensions. However, these examples may still be of interest in
physics because the theory of violent relaxation is valid for other
systems with long-range interactions described by the Vlasov equation
\cite{new}. The foregoing distributions may thus be relevant for
one-dimensional systems. They can also be
relevant in 2D turbulence (see Sec. \ref{sec_turb}) where the energy
$\epsilon=v^2/2+\Phi({\bf r})$ is replaced by the stream function
$\psi({\bf r})$, so that there is no condition of normalization
equivalent to $\int f d{\bf v}<\infty$. 

The $E^{-1}$ behaviour of $\overline{f}(E)$ for $E\rightarrow +\infty$
arises because we have assumed that the function $\chi(\eta)$ is
regular at $\eta=0$.  In fact, the level $\eta=0$ plays a particular
role in the theory because it corresponds to the ``vaccum'' which has
a very large phase space extension and which can mix with the non-zero
levels. Therefore, we expect that $\chi(\eta)\rightarrow \chi_{0}\delta(\eta)$
for $\eta\rightarrow 0$.  As a consequence, the level $\eta=0$  should be
treated specifically, and a more physical form of partition
function, which isolates the contribution of $\eta=0$, would be
\begin{equation}
\label{sn1}
Z(E)=1+\int_{a}^{+\infty} \chi(\eta)e^{-\eta E}d\eta,
\end{equation}
where $a\ge 0$. Note that we can take $\chi_{0}=1$ without restriction
of generality so that the value of $\gamma(0)$, which is infinite,
never appears in the theory. If we reconsider example (i) with now
$\chi=1/(b-a)$ for $a\le
\eta\le b$ and $\chi=0$ otherwise, we get
\begin{equation}
\label{sn2}
\overline{f}(E)={e^{-aE}-e^{-bE}+E(a e^{-aE}-b e^{-bE})\over E\lbrack E(b-a)+e^{-aE}-e^{-bE}\rbrack}.
\end{equation}
If $a\neq 0$ (gap), the DF decreases as $\overline{f}\sim
a/(b-a)E^{-1}e^{-aE}$ for $E\rightarrow +\infty$ and if $a = 0$ (no
gap) as $\overline{f}\sim (1/b)E^{-2}$. The density profile $\rho=\int
f d{\bf v}$ is now well-defined in $d=3$. If we reconsider example
(ii) with (\ref{sn1}), we get
\begin{equation}
\label{sn3}
\overline{f}(E)={cb\over (1+bE)\lbrack (1+bE)^{c}+1\rbrack},
\end{equation}
which decreases as $\overline{f}\sim c b^{-c}E^{-(c+1)}$. We think
that the particularity of the level $\eta=0$ is an important point
that deserves further consideration.

\subsection{Incomplete relaxation, Tsallis entropies and $H$-functions}
\label{sec_incomplete}

The statistical approach presented previously rests on the assumption
that the collisionless mixing is efficient so that the ergodic
hypothesis which sustains the statistical theory is fulfilled.  In
reality, this is not the case.  It has been understood since the
beginning \cite{lb} that violent relaxation is {\it incomplete} so
that the mixing entropy (\ref{E9}) is {\it not} maximized in the whole
phase space and real stellar systems are not described by
Lynden-Bell's statistics. In fact, for stellar systems, violent
relaxation cannot be complete because there is no maximum entropy
state in an unbounded domain. The generalized isothermal distribution
functions (\ref{E14}) predicted by Lynden-Bell, when coupled to the Poisson
equation, yield density profiles whose mass is infinite (the density
decreases as $r^{-2}$ at large distances). But this mathematical
problem is rather independent from the {\it physical} reason why
violent relaxation is incomplete. Physically, real stellar systems
tend towards the maximum entropy state during violent relaxation but
cannot attain it because the gravitational potential variations die
away before the relaxation process is complete. Thus, for {\it
dynamical} reasons, the system will not explore the whole phase space
ergodically as discussed in
\cite{csr,mn}. However, since the Vlasov equation admits an infinite
number of stationary solutions, the coarse-grained distribution
$\overline{f}$ can be trapped in one of them and remain frozen in that
state until collisional effects come into play (on longer
timescales). This steady solution is not, in general, the most mixed
state (it is only partially mixed) so it differs from Lynden-Bell's
statistical prediction. The concept of incomplete violent relaxation
explains why galaxies are more confined than predicted by statistical
mechanics (the density profile of elliptical galaxies decreases as
$r^{-4}$ instead of $r^{-2}$ \cite{bt}).

In order to quantify the importance of mixing, Tremaine, H\'enon \&
Lynden-Bell \cite{thlb} have introduced the notion of $H$-functions. They are
defined by
\begin{equation}
H[\overline{f}]=-\int C(\overline{f}) d^{3}{\bf r}d^{3}{\bf v}, \label{i1}
\end{equation}
where $C$ is any convex function. It can be shown that the
$H$-functions $H[\overline{f}]$ calculated with the coarse-grained
distribution function increase during violent relaxation in the sense
that $H[\overline{f}({\bf r},{\bf v},t)]\ge H[\overline{f}({\bf
r},{\bf v},0)]$ for $t>0$ where it is assumed that, initially, the
system is not mixed so that $\overline{f}({\bf r},{\bf v},0)={f}({\bf
r},{\bf v},0)$. This is similar to the $H$-theorem in kinetic theory.
However, contrary to the Boltzmann equation, the Vlasov equation does
not single out a unique functional (the above inequality is true for
all $H$-functions) and the time evolution of the $H$-functions is not
necessarily monotonic (nothing is implied concerning the relative
values of $H(t)$ and $H(t')$ for $t,t'>0$). Yet, this observation
suggests a notion of generalized {\it selective decay principle}:
among all invariants of the collisionless dynamics, the $H$-functions
(fragile constraints) tend to increase ($-H$ decrease) on the
coarse-grained scale while the mass and the energy (robust
constraints) are approximately conserved.  According to this
phenomenological principle, we {\it might} expect (see however the
last paragraph of this section) that the metaequilibrium state reached
by the system as a result of incomplete violent relaxation will
maximize a certain $H$-function (non-universal) at fixed mass and
energy.  Repeating the calculations of Sec. \ref{sec_ge} with
$H[\overline{f}]$ instead of $S[\overline{f}]$, the extremization of a
$H$-function at fixed $E$ and $M$ determines a distribution function
$\overline{f}=\overline{f}(\epsilon)$ with $\overline{f}'(\epsilon)<0$
which is a stationary solution of the Vlasov equation (recall that our
argument applies to the {\it coarse-grained} distribution).  Moreover,
if the DF {\it maximizes} the $H$-function at fixed $E$ and $M$, then
it is nonlinearly dynamically stable with respect to the
Vlasov-Poisson system
\cite{thlb,aa3,cst}\footnote{During mixing 
$D\overline{f}/Dt\neq 0$ and the $H$-functions $H[\overline{f}]$
increase. Once it has mixed $D\overline{f}/Dt= 0$ so that $\dot
H[\overline{f}]=0$. Since $\overline{f}({\bf r},{\bf v},t)$ has been
brought to a maximum $\overline{f}_{0}({\bf r},{\bf v})$ of a certain
$H$-function and since $H[\overline{f}]$ is conserved (after mixing),
then $\overline{f}_{0}$ is a nonlinearly dynamically stable steady
state of the Vlasov equation.}. In general, the $H$-function
$H^{*}[\overline{f}]$ that is effectively maximized by the system as a
result of incomplete violent relaxation (if any) is difficult to
predict \cite{gfp}.  It depends on the initial conditions (due to the
Casimirs) {\it and} on the efficiency of mixing. If mixing is complete
(as may be the case for systems others than gravitational ones), the
$H$-function that is maximized at equilibrium is the generalized
entropy (\ref{s3b}), hence $H^{*}[\overline{f}]=S[\overline{f}]$, and
the stationary distribution function is the Lynden-Bell distribution
(\ref{E13}). If mixing is incomplete, $H^{*}[\overline{f}]$ and
$\overline{f}(\epsilon)$ can take forms that are {\it not} compatible
with the expressions (\ref{s3b}) and (\ref{E13}) derived in the
statistical approach.

In the context of incomplete violent relaxation, the  Tsallis
functional
\begin{equation}
S_{q}[\overline{f}]=-{1\over q-1}\int (\overline{f}^{q}-\overline{f}) d^{3}{\bf r}d^{3}{\bf v},
\label{tsal1}
\end{equation}
is a particular $H$-function whose maximization at fixed mass and
energy leads to distribution functions of the form
\begin{equation}
\label{q} \overline{f}({\bf r},{\bf v})=\biggl\lbrack \mu-{\beta(q-1)\over
q}\epsilon\biggr\rbrack^{1\over q-1}.
\end{equation}
These distribution functions characterize stellar polytropes
\cite{bt}. They are particular stationary solutions of the Vlasov equation. 
For $q>1$, the polytropic distribution functions have a compact
support (they vanish at a maximum energy $\epsilon_{max}$) unlike the
Lynden-Bell distribution functions (\ref{E13}) whose tails extend to
infinity. Stellar polytropes with index $n\le 5$ (where
$n=3/2+1/(q-1)$) describe {confined} structures with finite mass,
unlike isothermal stellar systems. They have been studied for a long
time in astrophysics as simple mathematical models of stellar
systems. Unfortunately, pure polytropic distributions do not provide a
good model of incomplete violent relaxation for elliptical galaxies
\cite{bt}. An improved model is a {\it composite model} that is
isothermal in the core (justified by Lynden-Bell's theory of violent
relaxation) and polytropic in the halo (due to incomplete relaxation)
with an index $n=4$ \cite{hm,aa3}. Since the maximization principle
determining the nonlinear dynamical stability of a collisionless
stellar system (maximization of a $H$-function at fixed mass and
energy) is {\it similar} to the maximization principle determining the
thermodynamical stability of a collisional stellar system
(maximization of the Boltzmann entropy at fixed mass and energy) we
can use a {\it thermodynamical analogy} and develop an {\it effective
thermodynamical formalism} (E.T.F.) to analyze the nonlinear dynamical
stability of collisionless stellar systems \cite{gfp,aa3,cst}.  We
emphasize, however, that the maximization of a $H$-function at fixed
mass and energy is a condition of nonlinear dynamical stability for
the Vlasov equation, not a condition of thermodynamical
stability. Therefore, this thermodynamical analogy is purely
formal. In particular, in the context of violent relaxation, Tsallis
functional $S_{q}[\overline{f}]$ is a particular $H$-function, not an entropy.

If we were to apply Tsallis generalized thermodynamics in the context
of violent relaxation, we would need to replace the Lynden-Bell
entropy (\ref{E9}) by the $q$-entropy
\begin{equation}
S_{q}[\rho]=-{1\over q-1}\int (\rho^{q}({\bf r},{\bf
v},\eta)-\rho({\bf r},{\bf v},\eta))d^{3}{\bf r}d^{3}{\bf v}d\eta,
\label{se6}
\end{equation}
as argued in \cite{brands}. The generalized mixing entropy
$S_{q}[\rho]$, which is a functional of the {\it probability} $\rho({\bf
r},{\bf v},\eta)$, would be the proper form of $q$-entropy in that
context, taking into account the specificities of the collisionless
dynamics. For $q\rightarrow 1$, it returns the Lynden-Bell entropy
(\ref{E9}). For $q\neq 1$, it could take into account incomplete
mixing and non-ergodicity. In that context, the $q$ parameter could be
interpreted as a {\it measure of mixing} and Tsallis entropy could be
interpreted as a functional attempting to take into account
non-ergodicity in the process of incomplete violent
relaxation. Maximizing $S_{q}[\rho]$ at fixed mass, energy and
Casimirs, we obtain a $q$-generalization of the Gibbs state
(\ref{E11}). This maximization principle is a condition of
thermodynamical stability (in Tsallis generalized sense) in the
context of violent relaxation. Then, we can obtain a
$q$-generalization of the equilibrium coarse-grained distribution
function (\ref{E13}) in a fashion similar to that of
Sec. \ref{sec_meta}, after introducing proper averaging procedures
(e.g., $q$-expectation values). For appropriate values of $q$, these
distribution functions will have finite mass contrary to Lynden-Bell's
distribution. We shall not try, however, to develop this generalized
formalism in more detail here. Note that in the case of two levels
$f\in \lbrace 0,\eta_{0}\rbrace$, and in the dilute limit of the
theory $\overline{f}\ll\eta_0$, $S_{q}[\rho]$ can be written in terms
of the coarse-grained distribution $\overline{f}=\rho\eta_{0}$ in the
form $S_{q}[\overline{f}]=-{1\over q-1}\int \lbrack
(\overline{f}/\eta_0)^{q}-(\overline{f}/\eta_0)\rbrack d^{3}{\bf
r}d^{3}{\bf v}$. In this particular limit, Tsallis functional
$S_{q}[\overline{f}]$ could be interpreted as a generalized entropy
(not just a $H$-function). Therefore, Tsallis functional $S_{q}[\rho]$
expressed in terms of $\rho({\bf r},{\bf v},\eta)$ is a generalized
entropy while Tsallis functional $S_{q}[\overline{f}]$ expressed in
terms of $\overline{f}({\bf r},{\bf v})$ is either a $H$-function
(dynamics) or a particular case of entropy $S_{q}[\rho]$
(thermodynamics) for two levels in the dilute limit. However, it is
not clear why complicated effects of non-ergodicity (incomplete
mixing) could be encapsulated in a simple functional such as
(\ref{se6}). Indeed, other functionals of the form $S=-\int
C(\rho)d\eta d{\bf r}d{\bf v}$ where $C$ is convex could be considered
as well. {\it As discussed above, the observations of galaxies do not
support the prediction of non-extensive thermodynamics obtained by
maximizing Tsallis $q$-entropy (\ref{se6}).}  Furthermore, it is not
clear whether the idea of changing the form of entropy in case of
incomplete relaxation is the most relevant. An alternative approach
developed in
\cite{csr,mn} is to keep the Lynden-Bell entropy (\ref{E9}) unchanged 
but describe the dynamical evolution of $\rho({\bf r},{\bf v},t)$ 
by a relaxation equation of the form
\begin{equation}
{\partial\rho\over\partial t}+{\bf v}\cdot {\partial\rho\over\partial {\bf r}}-\nabla\Phi\cdot {\partial\rho\over\partial {\bf v}}={\partial\over\partial {\bf v}}\cdot \biggl\lbrace D({\bf r},{\bf v},t)\biggl\lbrack {\partial\rho\over\partial {\bf v}}+\beta(t)(\eta-\overline{f})\rho {\bf v}\biggr\rbrack\biggr\rbrace,
\label{relaxcgw}
\end{equation}
with a diffusion coefficient $D({\bf r},{\bf v},t)$ going to zero for
large time (as the variations of the gravitational potential $\Phi$
decay) and in regions of phase-space where the fluctuations
$\delta\Phi$ are not strong enough to provide efficient mixing. The
vanishing of the diffusion coefficient can ``freeze'' the system in a
subdomain of phase space and account for incomplete relaxation and
non-ergodicity. In general, the resulting state, although incompletely
mixed, is not a $q$-distribution. This approach in interesting because
it is not based on a generalized entropy, so there is no free
parameter like $q$ or $C(\rho)$. However, it demands to solve a
dynamical equation (\ref{relaxcgw}) to predict the equilibrium
state. {\it The idea is that, in case of incomplete relaxation
(non-ergodicity), the prediction of the equilibrium state is
impossible without considering the dynamics.}

We would like to emphasize again the distinction between entropies and
$H$-functions. An entropy is a quantity which is proportional to the
logarithm of the disorder, where the disorder is equal to the number
of microstates consistent with a given macrostate. This is how the
Lynden-Bell entropy (\ref{E9}) has been defined. Tsallis entropy
(\ref{se6}) could be considered as a generalization of this definition
in the case where the phase-space has a complex structure so that the
evolution is non-ergodic.  In each case, the entropy is a functional
of the probability $\rho({\bf r},{\bf v},\eta)$ and the maximization
of these entropies at fixed mass, energy and Casimirs is a condition
of thermodynamical stability.  The $H$-functions do not have a
statistical origin. They are just arbitrary functionals of the
coarse-grained distribution $\overline{f}({\bf r},{\bf v},t)$ of the
form (\ref{i1}). They are useful to characterize the degree of mixing
of a collisionless stellar system \cite{thlb}. Furthermore, their
maximization at fixed mass and energy provides a condition of
nonlinear dynamical stability with respect to the Vlasov
equation. Finally, the ``generalized entropies'' (\ref{g2}) defined in
Sec. \ref{sec_ge} can be regarded as entropies which are proportional
to the logarithm of the number of microstates consistent both with a
given macrostate and with the constraints imposed by the Vlasov
equation (Casimirs). Their functional form depends on the initial
condition. They are defined on a projection space
($\overline{f}$-space) where a macrostate is defined by the
specification of $\overline{f}({\bf r},{\bf v})$ instead of $\rho({\bf
r},{\bf v},\eta)$.

Finally, we note that the maximization of the Lynden-Bell entropy
(\ref{E9}), of the Tsallis entropy (\ref{se6}) or of a H-function
(\ref{i1}) leads to a distribution function of the form
$\overline{f}=\overline{f}(\epsilon)$ with $\overline{f}'(\epsilon)<0$
depending only on the energy. These DF  can only describe {\it spherical}
stellar systems (and even a sub-class of them) \cite{bt}. In reality,
stellar systems are not spherical and their distribution functions are
not function of the energy alone. Indeed, according to the Jeans
theorem
\cite{bt}, there exists more general stationary solutions of the
Vlasov equation which depend on other integrals of motion. This
indicates that the structure of the final state of a collisionless
stellar system depends on its dynamical evolution in a complicated
manner. An important problem in astrophysics is therefore to find the
form of distribution function appropriate to real galaxies. Simple
concepts based on entropies and $H$-functions are not sufficient to
understand the structure of galaxies. This is particularly
deceptive. However, conceptually, the theory of violent relaxation is
important to explain {\it how} a collisionless stellar system reaches
a steady state. This is due to phase mixing in phase space. The
coarse-grained DF $\overline{f}({\bf r},{\bf v},t)$ reaches a steady
state $\overline{f}({\bf r},{\bf v})$ in a few dynamical times while
the fine-grained distribution function ${f}({\bf r},{\bf v},t)$
develops filaments at smaller and smaller scales and is never steady
(presumably). Since this mixing process is very complex, the resulting
structure $\overline{f}({\bf r},{\bf v})$ should be extremely robust
and should be therefore a nonlinearly dynamically stable stationary
solution of the Vlasov equation. {\it Thus, the theory of incomplete
violent relaxation explains how collisionless stellar systems can be
trapped in nonlinearly dynamically stable stationary solutions of the
Vlasov equation on the coarse-grained scale}.

\section{Two-dimensional turbulence}
\label{sec_turb}

\subsection{Statistical mechanics of 2D vortices}
\label{sec_gs}

The same ideas apply in 2D turbulence to understand the formation of
coherent structures (jets and vortices) in large-scale flows. The
analogy between stellar systems and 2D vortices is discussed in
Chavanis \cite{houches}. A statistical theory of point vortices has
been first developed by Onsager \cite{onsager} and Joyce \& Montgomery
\cite{jm}. This theory predicts the statistical
equilibrium state of a point vortex gas, reached for $t\rightarrow
+\infty$ after a ``collisional'' relaxation, assuming ergodicity. The
most probable vorticity profile is given by
\begin{equation}
\omega({\bf r})=-\Delta\psi=\sum_{i}A_{i}e^{-\beta\gamma_{i}\psi},
\label{gs1}
\end{equation}
which is similar to the statistical distribution (\ref{bol10}) of a
multi-components system of stars (note that the vorticity is
proportional to the density of point vortices). A kinetic theory of
point vortices has been developed by Dubin \& O'Neil \cite{dn} and
Chavanis \cite{pvkin,houches}. The collision term of the derived
kinetic equation, which is the counterpart of the Landau equation
(\ref{bol13}), cancels out when the profile of angular momentum is
monotonic so that this equation (valid to order $1/N$) does not relax
towards the statistical equilibrium state. This implies that the
relaxation time scale (if there is ever relaxation) is larger than $N
t_{D}$.

In the limit $N\rightarrow +\infty$, the evolution of the
system is described by the 2D Euler equation (\ref{ha3}) which is the
counterpart of the Vlasov equation (\ref{vp1}). The statistical
mechanics of continuous vorticity fields described by the 2D Euler
equation has been developed by Miller and Robert \& Sommeria
\cite{miller}. This is similar to the theory of violent relaxation of
Lynden-Bell \cite{lb,csr}. In that context, we speak of ``inviscid
relaxation'' or ``chaotic mixing''. The mixing entropy is
\begin{equation}
S[\rho]=-\int \rho({\bf r},\sigma)\ln\rho({\bf r},\sigma) d^{2}{\bf r}d\sigma,
\label{gs2a}
\end{equation}
and the Gibbs state reads
\begin{equation}
\rho({\bf r},\sigma)={1\over Z}\chi(\sigma)e^{-\sigma(\beta\psi+\alpha)},
\label{gs2}
\end{equation}
with notations similar to those of Sec. \ref{sec_meta} (here, $\sigma$
labels the vorticity levels).  The density probability $\rho({\bf
r},\sigma)$ gives the local distribution of vorticity at statistical
equilibrium. It maximizes the mixing entropy (\ref{gs2a}) at fixed
energy $E={1\over 2}\int \overline{\omega}\psi d^{2}{\bf r}$, circulation
$\Gamma=\int \overline{\omega}d^{2}{\bf r}$ (robust constraints) and
Casimir constraints or fine-grained moments $\Gamma^{f.g.}_{n>1}=\int
\overline{\omega^{n}}d^{2}{\bf r}=\int \rho\sigma^{n}d\sigma d^{2}{\bf r}$
(fragile constraints). The partition function can be written
\begin{equation}
Z=\int_{-\infty}^{+\infty} \chi(\sigma)e^{-\sigma(\beta\psi+\alpha)}d\sigma,
\label{gs3}
\end{equation}
and the most probable coarse-grained vorticity
$\overline{\omega}=\int \rho\sigma d\sigma$ is related to the
stream function by a relation of the form
\begin{equation}
\overline{\omega}=-{1\over\beta}{\partial \ln Z\over\partial \psi}=F(\beta\psi+\alpha)=f(\psi).
\label{gs4}
\end{equation}
This is a steady state of the 2D Euler equation where $f$ is monotonic
(since $f'(\psi)=-\beta\omega_{2}$ with
$\omega_{2}=\overline{\omega^{2}}-\overline{\omega}^{2}\ge 0$, it is
increasing at negative temperatures and decreasing at positive
temperatures).  Note that the vorticity levels $\sigma$ can take
positive and negative values contrary to the case of self-gravitating
systems for which $\eta\ge 0$. Note also that ${\omega}$ is a
vorticity field not a distribution of particles, unlike ${f}$ in
astrophysics (only in the point vortex model can we interprete $\omega$
as a distribution of particles since it is related to the density of
point vortices). The most probable coarse-grained vorticity
(\ref{gs4}) maximizes a generalized entropy
\begin{equation}
S[\overline{\omega}]=-\int C(\overline{\omega})d^{2}{\bf r},
\label{gs5}
\end{equation}
at fixed circulation and energy. Indeed, this optimization problem leads
to a relation of the form
\begin{equation}
C'(\overline{\omega})=-\beta\psi-\alpha,
\label{gs6}
\end{equation}
which can be identified with Eq. (\ref{gs4}) with
$f'(\psi)=-\beta/C''(\overline{\omega})$. This identification relates
the function $C(\overline{\omega})$ to the function $F(x)$ whose form
depends on $\chi(\sigma)$ through Eqs. (\ref{gs3}) and
(\ref{gs4}). Explicitly, we have
\begin{equation}
C(\overline{\omega})=-\int^{\overline{\omega}} F^{-1}(x)dx.
\label{gs7b}
\end{equation}

We can also introduce a notion of generalized {\it selective decay
principle} in 2D turbulence: among all inviscid invariants of the 2D
Euler equation, the $H$-functions (fragile constraints)
$H[\overline{\omega}]=-\int C(\overline{\omega})d^{2}{\bf r}$ increase
($-H$ decrease) on the coarse-grained scale or in the presence of a
small viscosity (Appendix \ref{app_h}) while the energy
$E[\overline{\omega}]$ and the circulation $\Gamma[\overline{\omega}]$
(robust constraints) are approximately conserved.  Therefore, the
metaequilibrium state resulting from violent relaxation is expected to
maximize a certain $H$-function (non-universal) at fixed energy and
circulation.  This generalizes the usual {selective decay principle}
of 2D turbulence which considers the minimization of enstrophy
$\Gamma_2=\int \omega^2 d^{2}{\bf r}$ at fixed energy and
circulation. In our approach, minus the enstrophy
$-\Gamma_2[\overline{\omega}]=-\int \overline{\omega}^{2}d^{2}{\bf r}$
and the Tsallis functionals $S_q[\overline{\omega}]=-{1\over q-1}\int
(\overline{\omega}^q-\overline{\omega})d^{2}{\bf r}$ are particular
$H$-functions (note that the enstrophy $\Gamma_{2}$ is a particular
case of Tsallis functional with $q=2$).

The extremization of a $H$-function at fixed energy and circulation
leads to a stationary solution of the 2D Euler equation of the form
$\overline{\omega}=f(\psi)$ where $f$ is a monotonic function
specified by the convex function $C(\omega)$. Furthermore, as shown in
\cite{ellis}, the condition of {\it maximum}  provides a refined criterion of nonlinear dynamical
stability for the 2D Euler-Poisson system (the physical interpretation
of this criterion applying to the {\it coarse-grained} vorticity is
the same as in the remark of Sec. \ref{sec_incomplete}).  Note that
contrary to the Vlasov equation, the relation $\omega=f(\psi)$ is the
general form of stationary solution of the 2D Euler equation (for
systems with no special symmetries). Therefore, in 2D hydrodynamics,
{\it any} nonlinearly dynamically stable stationary solution of the 2D
Euler equation maximizes a $H$-function at fixed circulation and
energy (and, possibly, angular momentum and impulse) contrary to the
case of the Vlasov equation in astrophysics where a more general class
of steady solutions exists due to the Jeans theorem.

Finally, the Tsallis entropy in the context of the 2D Euler equation
is a functional of the vorticity distribution $\rho({\bf r},\sigma)$
of the form $S_{q}[\rho]=-{1\over q-1}\int (\rho^{q}-\rho)d\sigma
d^{2}{\bf r}$ generalizing the mixing entropy (\ref{gs2a})
\cite{brands}. This functional could be an attempt to take into
account non-ergodicity in the process of violent relaxation of 2D
turbulent flows. However, other functionals could be considered as
well, and Tsallis entropy does not provide a correct description of
non-ergodicity in all observed cases. This means that the {\it type of
mixing} in 2D turbulence (and stellar dynamics) is more complex than
the one (multi-fractal) described by the Tsallis functional
\cite{tsallis}. Non-ergodicity (incomplete relaxation) can be taken 
into account dynamically by using relaxation equations with a space
dependent diffusion coefficient related to the fluctuations \cite{rsprl,csr}.

\subsection{Prior vorticity distribution}
\label{sec_prior}

The statistical approach of Miller-Robert-Sommeria applies to flows
that are strictly described by the 2D Euler equation. In this point of
view, one must conserve the value of all the Casimir invariants (or
vorticity moments). This leads to the expression (\ref{gs2}) for the
most probable distribution of vorticity, where the function
$\chi(\sigma)$ is determined by the initial conditions through the
value of the Casimir integrals (this is precisely the Lagrange
multiplier associated to these constraints). However, in geophysics,
there exists situations in which the flow is continuously forced at
small-scales so that the conservation of the Casimirs is
destroyed. Ellis {\it et al.} \cite{ellis} have proposed to take into
account these situations by fixing the function $\chi(\sigma)$ instead
of the Casimirs. Physically, this prior vorticity distribution can be
viewed as a {\it global} distribution of vorticity imposed by a
small-scale forcing. It can be due to convection and 3D effects like
in the atmosphere of Jupiter. Its specific form has to be adapted to
the situation.  Then, two-dimensional turbulence organizes this global
distribution of vorticity into large-scale coherent structures.  These
organized states result from a balance between entropic and
energetic effects: the system tends to mix but complete mixing, which
would result in a uniform distribution, is prevented by the energy
constraint.  The most probable {\it local} distribution of vorticity
is now obtained by maximizing a relative entropy conditioned by the
prior distribution
\begin{equation}
S_{\chi}[\rho]=-\int \rho({\bf r},\sigma)\ln\biggl\lbrack {\rho({\bf
r},\sigma)\over \chi(\sigma)}\biggr\rbrack d^{2}{\bf r}d\sigma,
\label{rel}
\end{equation}
at fixed circulation and energy (no other constraints). The
conservation of the Casimirs has been replaced by the specification of
a prior distribution $\chi(\sigma)$. As shown in Chavanis \cite{geo},
the relative entropy (\ref{rel}) can be seen as a Legendre transform
$S_{\chi}=S-\sum_{n>1}\alpha_{n}\Gamma_{n}^{f.g.}$ of the mixing
entropy (\ref{gs2a}) when the constraints associated with the
conservation of the vorticity moments (Casimirs) are treated
canonically. Indeed, the approach of Ellis {\it et al.}
\cite{ellis} amounts to fixing the conjugate variables
$\alpha_{n>1}$ instead of the fine-grained moments
$\Gamma_{n>1}^{f.g.}$. If we view the vorticity levels as species of
particles, this is equivalent to fixing the chemical potentials
instead of the total number of particles in each species. This assumes
that the 2D system is in contact with a sort of ``reservoir''. The
forcing and dissipation break the conservation of the Casimirs and
impose instead a distribution of vorticity. By contrast, the robust
constraints (circulation and energy) are still treated
microcanonically. The maximization of $S_{\chi}$ at fixed $E$,
$\Gamma$ again leads to the distribution (\ref{gs2}) but with a
different interpretation. In the present context, the statistical
equilibrium state results from an interplay between 3D effects (the
non-universal small-scale homogeneous forcing encapsulated in the
prior $\chi(\sigma)$) and 2D effects (the universal Gibbs factor
$e^{-\sigma(\alpha+\beta\psi)}$ giving rise to inhomogeneous
large-scale structures). The statistical distribution is the product
of these two effects. The partition function and the most probable
coarse-grained vorticity field are still given by Eqs. (\ref{gs3}) and
(\ref{gs4}). However, in this new approach, the
function $F(x)$ is fixed directly by the prior vorticity distribution
$\chi(\sigma)$ while in the approach of Miller-Robert-Sommeria, it has
to be related a posteriori to the initial conditions in a 
complicated way.

The approach of Ellis {\it et al.} \cite{ellis} is very close to the
notion of superstatistics since it considers that the fluctuations of
vorticity $\chi(\sigma)$ are given {\it a priori} by an external process,
which is also the case for the fluctuations of temperature $f(\beta)$
in the Beck-Cohen superstatistics. Therefore, the $\omega-\psi$
relationship and the generalized entropy $S[\overline{\omega}]$ are
directly determined by the prior vorticity distribution $\chi(\sigma)$
through the formula
\begin{equation}
\label{s3c} C(\overline{\omega})=-\int^{\overline{\omega}} \lbrack (\ln
\hat\chi)'\rbrack^{-1}(-x)dx,
\end{equation}
where
$\hat\chi(E)=\int_{-\infty}^{+\infty}\chi(\sigma)e^{-\sigma E}d\sigma$,
according to Eqs. (\ref{gs7b}), (\ref{gs4}) and (\ref{gs3}). This
makes the generalized entropy $S[\overline{\omega}]$ an {\it
intrinsic} quantity. In the present context, it is determined by the
small-scale forcing (through the prior $\chi$) while in the approach
of Miller-Robert-Sommeria it depends on the initial conditions
(through the Casimirs).  Furthermore, in the present context,
$S[\overline{\omega}]$ really has the status of an entropy in the
sense of the large deviation theory. Indeed, Ellis {\it et al.} 
\cite{ellis} show that the probability of the coarse-grained vorticity
field $\overline{\omega}({\bf r})$ at statistical equilibrium can be
written in the form of the Cramer formula
\begin{equation}
\label{cram}P[\overline{\omega}]\sim e^{nS[\overline{\omega}]},
\end{equation}
where $n$ is the number of sites of the underlying lattice introduced
in their mathematical analysis. Therefore, the most probable vorticity
field $\overline{\omega}$ maximizes $S[\overline{\omega}]$ at fixed
circulation and energy. This maximization principle also provides a
refined condition of nonlinear dynamical stability with respect to the
2D Euler-Poisson system \cite{ellis}.

\subsection{Example of generalized entropy}
\label{sec_ex}

Let us consider, for illustration, the prior vorticity
distribution $\chi(\sigma)$ introduced by Ellis et al.
\cite{ellis} in their model of jovian vortices. It corresponds to a
de-centered Gamma distribution
\begin{equation}
\chi(\sigma)={1\over |\epsilon|}R\biggl \lbrack {1\over\epsilon}\biggl (\sigma+{1\over\epsilon}\biggr );{1\over\epsilon^{2}}\biggr \rbrack,
\label{e1}
\end{equation}
where $R(z;a)=\Gamma(a)^{-1}z^{a-1}e^{-z}$ for $z\ge 0$ and $R=0$
otherwise. The scaling of $\chi(\sigma)$ is chosen such that
$\langle\sigma\rangle=0$, ${\rm var}(\sigma)=1$ and ${\rm
skew}(\sigma)=2\epsilon$. This distribution is a variant of Gamma
distribution considered by Beck \& Cohen \cite{bc}. Setting
$E\equiv \beta\psi+\alpha$, we get
\begin{equation}
Z(E)=\hat\chi(E)={e^{E/\epsilon}\over (1+\epsilon
E)^{1/\epsilon^{2}}}, \label{e2}
\end{equation}
and
\begin{equation}
\overline{\omega}(E)=-(\ln Z)'(E)={-E\over 1+\epsilon E}.
\label{e3}
\end{equation}
Inversing the relation (\ref{e3}), we obtain
\begin{equation}
\label{e4}
-E={\overline{\omega}\over 1+\epsilon\overline{\omega}}=C'(\overline{\omega}).
\end{equation}
After integration, we obtain the generalized entropy
\begin{equation}
\label{e5}
C(\overline{\omega})={1\over\epsilon}\biggl\lbrack \overline{\omega}-{1\over\epsilon}\ln (1+\epsilon\overline{\omega})\biggr\rbrack.
\end{equation}
This form of entropy can also be obtained from the techniques of
the large deviation theory as discussed in \cite{ellis}. Our approach,
leading to the general formula (\ref{s3c}), is a simple
alternative to obtain the generalized entropy
$C(\overline{\omega})$ associated to the prior vorticity
distribution $\chi(\sigma)$. On the other hand, for a Gaussian prior distribution 
\begin{equation}
\chi(\sigma)=e^{-{\sigma^{2}\over 2}},
\label{gg1}
\end{equation}
we get
\begin{equation}
Z(E)=\sqrt{2\pi} e^{E^{2}\over 2}, \qquad 
\overline{\omega}(E)=-E, \qquad C(\overline{\omega})={1\over 2}\overline{\omega}^{2}.
\label{gg2}
\end{equation}
Therefore, the $\overline{\omega}-\psi$ relationship is linear and the generalized entropy  $S[\overline{\omega}]=-{1\over 2}\int
\overline{\omega}^{2}d^{2}{\bf r}$ is minus the enstrophy. It also corresponds to the limit of Eq. (\ref{e5}) for $\epsilon\rightarrow 0$.
 Other examples of prior vorticity distributions are collected in \cite{gfp}.
An example which has not been given previously is when $\chi(E)$ is of the Tsallis form
\begin{equation}
\chi(\sigma)=\biggl (1-{1\over 2p}\sigma^{2}\biggr )^{p}, \qquad |\sigma|\le \sqrt{2p}.
\label{gg3}
\end{equation}
For $p\rightarrow +\infty$, we recover the Gaussian distribution
(\ref{gg1}). For the distribution (\ref{gg3}), we get
\begin{equation}
Z(E)=2^{(3+2p)/4}p^{(5-2p)/4}\sqrt{\pi}\Gamma(p)|E|^{-1/2-p}I_{1/2+p}(\sqrt{2p}|E|).
\label{gg4}
\end{equation}

\subsection{Generalized Fokker-Planck equations}
\label{sec_ax}

In the context of freely evolving 2D turbulence, a thermodynamical
parametrization of the 2D Euler equation has been proposed by Robert
\& Sommeria \cite{rsprl} in terms of relaxation equations based on a
maximum entropy production principle (MEPP). These equations conserve all the
Casimirs, increase the mixing entropy (\ref{gs2a}) and relax towards
the Gibbs state (\ref{gs2}). In the situations considered by Ellis et
al. \cite{ellis} where the system is forced at small scale, we have
proposed in \cite{geo} an alternative parametrization of the 2D Euler
equation. In that case, we have seen that only the energy and the
circulation (robust constraints) are conserved. The conservation of
the Casimirs is replaced by the specification of a prior vorticity
distribution $\chi(\sigma)$ encoding the small-scale forcing. This
fixes a form of generalized entropy (\ref{gs5}) through the formula
(\ref{s3c}). In that case, we have proposed to describe the
large-scale evolution of the flow on the coarse-grained scale by a
relaxation equation which conserves energy and circulation and
increases the generalized entropy (\ref{gs5}) until the equilibrium
state (\ref{gs4}) is reached. This can be obtained by using a
generalized Maximum Entropy Production Principle. The resulting
relaxation equation, introduced in
\cite{gfp}, has the form of a generalized Fokker-Planck equation
\begin{equation}
\label{ax1}
{\partial {\omega}\over\partial t}+{\bf u}\cdot \nabla {\omega}=\nabla\cdot \biggl\lbrace D\biggl\lbrack \nabla{\omega}+{\beta(t)\over C''({\omega})}\nabla\psi\biggr\rbrack\biggr\rbrace,
\end{equation}
\begin{equation}
\label{ax2}
\beta(t)=-{\int D\nabla{\omega}\cdot \nabla\psi d^{2}{\bf r}\over \int D{(\nabla\psi)^{2}\over C''({\omega})}d^{2}{\bf r}},
\end{equation}
where the evolution of the Lagrange multiplier $\beta(t)$ accounts for
the conservation of energy. Furthermore, the diffusion coefficient can
be obtained from a kinetic model leading to $D={K\epsilon^{2}/
\sqrt{C''(\omega)}}$ where $\epsilon$ is the resolution scale and
$K$ is a constant of order unity \cite{geo}. In these equations, the function
$C(\omega)$ is fixed by the prior distribution $\chi(\sigma)$.  These
equations are expected to be valid close to the equilibrium state in
the spirit of Onsager's linear thermodynamics. However, they may offer a
useful parametrization of 2D flows even if we are far from
equilibrium. Alternatively, according to the refined nonlinear
dynamical stability criterion of Ellis {\it et al.} \cite{ellis} these
relaxation equations can be used as powerful numerical algorithms to
compute arbitrary nonlinearly dynamically stable stationary solutions
of the 2D Euler-Poisson system. These ideas are further discussed in
\cite{geo} in relation with geophysical flows. We note that forced 2D
turbulence provides a physical situation of interest in which a
rigorous notion of generalized thermodynamics and generalized kinetics
emerges. In our formalism, all the complexity of the system is
encapsulated in a prior distribution $\chi(\sigma)$.  We can then
determine the generalized entropy $S[\overline{\omega}]$ by using 
formula (\ref{s3c}) and  substitute the result in the relaxation
equation (\ref{ax1}) to obtain the dynamical evolution of the
coarse-grained flow. The problem now amounts to finding the relevant
prior $\chi(\sigma)$. Of course, this depends
on the situation contemplated. Furthermore, for a given situation, it
is likely that a whole ``class'' of priors (or generalized entropies)
will sensibly give the same results. In practice, one 
has to proceed by trying and errors to find the relevant ``class of
equivalence'' adapted to the situation considered \cite{gfp}.

As discussed previously, the prior $\chi(\sigma)$ encodes the
small-scale forcing. It is due, e.g., to convection (in the jovian
atmosphere) or any other complicated process specific to the
situation contemplated. It is not our goal here to develop a precise model of
convection to determine a relevant form for $\chi(\sigma)$. We shall
rather remain at a phenomenological level and propose to describe the
generation of vorticity fluctuations by general stochastic
processes. Since the generating process must include a forcing and a
dissipation, we consider a generalized Langevin equation of the form introduced in \cite{gfp}:
\begin{equation}
\label{lan1} {d\sigma\over dt}=-\xi\sigma+\sqrt{2D\chi\biggl\lbrack {{\cal C}(\chi)\over \chi}\biggr\rbrack'}\eta(t),
\end{equation}
where $\eta(t)$ is a white noise and ${\cal C}(\chi)$ a convex
function of the global distribution of vorticity. The corresponding
(generalized) Fokker-Planck equation is
\begin{equation}
\label{lan2} {\partial\chi\over\partial
t}={\partial\over\partial\sigma}\biggl\lbrack D\chi {\cal
C}''(\chi){\partial\chi\over\partial
\sigma}+\xi\chi\sigma\biggr\rbrack.
\end{equation}
Its stationary solution determines the prior vorticity
distribution $\chi(\sigma)$ through the relation
\begin{equation}
\label{lan3} {\cal C}'(\chi)=-b{\sigma^2\over 2}-a,
\end{equation}
where $b=\xi/D$ is a sort of inverse temperature. For example, when
the coefficients of dissipation and forcing are constant,
corresponding to ${\cal C}(\chi)=\chi\ln\chi$ and leading to standard
stochastic processes, the prior distribution is the Gaussian
(\ref{gg1}) leading to a generalized entropy having the form of minus
the enstrophy $S[\overline{\omega}]=-\Gamma_{2}[\overline{\omega}]$
and to a linear $\overline{\omega}-\psi$ relationship at
equilibrium. However, our formalism allows to treat more general
situations. Furthermore, in the preceding discussion, we have
implicitly assumed that the prior relaxes more rapidly to its
equilibrium value than the coarse-grained vorticity field, so that, in
Eq. (\ref{ax1}), the generalized entropy $C(\omega)$ is calculated
from $\chi(\sigma)=\chi(\sigma,+\infty)$. This is probably a relevant
approximation. Otherwise, we need to couple the two equations
(\ref{ax1}) and (\ref{lan2}) and determine, at each time, the function
$C(\omega,t)$ from the prior $\chi(\sigma,t)$, using formula
(\ref{s3c}).

\section{Conclusion}
\label{conclusion}

In this paper, we have discussed some analogies between coarse-grained
distribution functions characterizing statistical equilibrium states
of collisionless stellar systems or inviscid 2D flows and the notion
of superstatistics introduced by Beck \& Cohen (2003). In particular,
we have shown that the coarse-grained distribution functions arising
in theories of violent relaxation can be viewed as forms of
superstatistics (albeit different from the Beck-Cohen
superstatistics). Although the concept of violent relaxation has been
introduced by Lynden-Bell (1967) long ago, it remains largely unknown
in the statistical mechanics community and this is why we have exposed
this theory in some detail here. Non-standard distributions arise on
the coarse-grained scale because they are expressed as averages of
fine-grained distributions. The observed (coarse-grained) distribution
function appears to be a superposition of Boltzmann's factors weighted
by a non-universal function $\chi(\eta)$ or $\chi(\sigma)$. To each
coarse-grained distribution, we can associate a generalized
entropy. For freely evolving systems, the functions $\chi(\eta)$ or
$\chi(\sigma)$ and the generalized entropies $S[\overline{f}]$ or
$S[\overline{\omega}]$ depend on the initial
conditions. Alternatively, in certain occasions, it may be justified
to regard the function $\chi$ as {\it imposed} by some external
processes. This prior distribution then directly determines the
generalized entropy.  This approach is particularly relevant in the
case of geophysical flows that are forced at small scales
\cite{ellis,geo}. It may also be valid in the case of dark matter
models in astrophysics were a small-scale forcing can alter the
conservation of the Casimirs and impose instead a distribution of
fluctuations. In these cases, the relaxation of the coarse-grained
field can be described by generalized Fokker-Planck equations where
the entropy is determined by the prior $\chi$
\cite{gfp,geo}. Alternatively, these relaxation equations can be used
as numerical algorithms to construct arbitrary nonlinearly dynamically
stable stationary solutions of the Vlasov and Euler equations
specified by a convex function $C$.

We have also discussed the two successive equilibrium states achieved
by a stellar system. In a first regime, the evolution is collisionless
and the system reaches a metaequilibrium state as a result of violent
relaxation. This is a nonlinearly dynamically stable stationary
solution of the Vlasov-Poisson system. On longer timescales, stellar
encounters (``collisions'') drive the system towards the statistical
equilibrium state described by the Boltzmann distribution (when the
escape of stars and the gravothermal catastrophe are prevented). The
metaequilibrium state (collisionless regime) and the statistical
equilibrium state (collisional regime) correspond to quite different
processes. They can be written as a superposition of Boltzmann factors
for each species of particles (collisional equilibrium) or for the
different phase levels (collisionless equilibrium).

In fact, violent relaxation is incomplete in general. A
famous example of incomplete relaxation in 2D turbulence is provided
by the plasma experiment of Huang \& Driscoll \cite{hd}. In this
experiment, the metaequilibrium state resulting from violent
relaxation has the form of a self-confined vortex surrounded by
un-mixed flow. This strong confinement is in contradiction with the
statistical mechanics of Miller-Robert-Sommeria
\cite{miller} which leads to un-restricted vorticity profiles. 
As discussed in Brands {\it et al.} \cite{brands}, the observed
confinement is due to incomplete relaxation and lack of
mixing/ergodicity. The system has evolved to a stationary solution of
the 2D Euler equation which is not the most mixed state. Now, any
nonlinearly dynamically stable stationary solution of the 2D Euler
equation maximizes a $H$-function $S[\overline{\omega}]$ at fixed
circulation and energy.  In the special case considered by Huang \&
Driscoll, this $H$-function turns out to be related to the enstrophy
functional $\Gamma_{2}[\overline{\omega}]$, which is a particular form
of the {\it Tsallis $H$-function} $S_{q}[\overline{\omega}]$ with $q=2$. This
``dynamical interpretation'' based on $H$-functions is different from
the ``generalized thermodynamical interpretation'' of Boghosian
\cite{boghosian} where $S_{q}[\overline{\omega}]$ is viewed 
as a {\it Tsallis $q$-entropy}. Since (in our sense) the Tsallis
functional $S_{q}[\overline{\omega}]$ is a $H$-function, not an
entropy, the use of $q$-expectation values is irrelevant in this
dynamical context. If we want to apply Tsallis thermodynamics in the
context of the 2D Euler equation, we need to introduce an entropy
$S_{q}[\rho]$ which is a functional of the {\it probability density}
$\rho({\bf r},\sigma)$.  However, in that case, the agreement with the
plasma experiment fails as shown in
\cite{brands}. Therefore, the experimental result of Huang \& Driscoll 
{\it cannot} in fact be explained by Tsallis generalized
thermodynamics when the full constraints of the Euler equation are
accounted for. The fact that the $\overline{\omega}-\psi$ relationship
resembles a $q$-distribution (in $\overline{\omega}$-space) is {\it
coincidental}. This is a particular solution of the Euler equation
resulting from incomplete violent relaxation. Since the 2D Euler
equation admits an infinity of stationary solutions, there are many
other examples of incomplete violent relaxation in 2D turbulence (and
stellar dynamics) where the system settles in a steady state that is
{\it not} described by the Tsallis distribution (in
$\overline{\omega}$-space or in $\rho$-space). The situation
described by Huang \& Driscoll in which an $\overline{\omega}-\psi$
relationship resembling a $q$-distribution emerges is fortuitous and
not generic.

In this paper, we have tried to distinguish different notions of
entropy that arise in the theory of violent relaxation. The {\it
mixing entropy} (\ref{E9})(\ref{gs2a}) is the fundamental entropy of
the theory.  It can be obtained by a combinatorial analysis and its
maximization at fixed mass/circulation, energy and Casimir invariants
determines the most probable distribution of fine-grained levels
$\rho({\bf r},{\bf v},\eta)$ through the Gibbs state
(\ref{E11})(\ref{gs2}), assuming ergodicity (complete mixing). The
{\it generalized mixing entropy} (\ref{se6}) is the appropriate
Tsallis generalization of (\ref{E9}) in the context of violent
relaxation. It can be seen as an attempt to take into account
non-ergodic effects and describe them in terms of a single parameter
$q$. All the machinery of non-extensive thermodynamics
($q$-expectation values,...) could be developed in that framework,
working with $\rho({\bf r},{\bf v},\eta)$ instead of $f({\bf r},{\bf
v})$. We might also consider other generalizations of entropy $S=-\int
C(\rho)d^{3}{\bf r}d^{3}{\bf v}d\eta$ where $C$ is convex. The status
of such generalizations is still in debate for the moment because it
is not clear whether non-ergodic effects can be encapsulated in a
simple functional. One must rather accept that the final state of the
system is unpredictable in case of incomplete violent relaxation. The
{\it relative entropy} (\ref{rel}) is the Legendre transform of the
mixing entropy (\ref{gs2a}) conditioned by a {prior} vorticity
distribution $\chi(\sigma)$ in the sense of
\cite{ellis,geo}. This description can be relevant for 2D turbulent
flows that are forced at small-scales. Its maximization at fixed
circulation and energy (no other invariants) determines the most
probable distribution of fine-grained levels $\rho({\bf r},\sigma)$
through the Gibbs state (\ref{gs2}) conditioned by an {\it imposed}
global distribution $\chi(\sigma)$. The {\it generalized entropy}
(\ref{g2})-(\ref{g8}) or (\ref{gs5})-(\ref{gs7b}) is the functional
that the most probable coarse-grained distribution $\overline{f}({\bf
r},{\bf v})$ or $\overline{\omega}({\bf r})$ given by
(\ref{E14})-(\ref{gs4}) maximizes at fixed energy and
mass/circulation. For freely evolving systems, it depends on the
initial conditions. For forced systems, it is determined by the prior
vorticity distribution $\chi(\sigma)$ through the formula (\ref{s3c}).
The {\it $H$-functions} (\ref{i1}) are arbitrary functionals (not
entropies) of the coarse-grained field. They increase during mixing
and their maximization at fixed mass/circulation and energy determines
a nonlinearly dynamically stable stationary solution of the
Vlasov/Euler equation with a monotonic relationship $f=f(\epsilon)$ or
$\omega=\omega(\psi)$. These stationary solutions can result from
complete or incomplete violent relaxation (in that case, $f$ and
$\omega$ must be regarded as the coarse-grained fields). When mixing
is complete, the $H$-function that is maximized at equilibrium is the
generalized entropy (\ref{g2})-(\ref{g8}) or
(\ref{gs5})-(\ref{gs7b}). When mixing is incomplete, the $H$-functions
and the coarse-grained distributions can take forms that are not
consistent with the statistical theory. For example, Tsallis
functional (\ref{tsal1}) is a particular $H$-function associated with
stellar polytropes and polytropic vortices. They form simple families
of stationary solutions of the Vlasov and 2D Euler equations. They
sometimes arise as a result of incomplete violent relaxation due to
the combined effect of Casimir constraints and non-ergodicity
\cite{hd,brands}. The maximization of a $H$-function at fixed
mass/circulation and energy is a condition of nonlinear dynamical
stability. We can develop a thermodynamical analogy and an effective
thermodynamical formalism to study the nonlinear dynamical stability
of the system, but the notion of ``generalized thermodynamics'' is
essentially effective in that context \cite{gfp,cst}.

In conclusion, a striking property of systems with long-range
interactions is the rapid emergence of coherent structures: galaxies
in astrophysics, vortices and jets in 2D turbulence, quasi-equilibrium
states in the HMF model... Since these metaequilibrium states are {\it
not} described by the Boltzmann distribution, some authors have
proposed to replace the Boltzmann entropy $S_{B}[f]$ by the Tsallis
entropy $S_{q}[f]$, invoking that the system is non-extensive so that
standard statistical mechanics is not applicable
\cite{boghosian,taruya,latora}. However, this approach ignores the
importance of the Vlasov equation and the concept of violent
relaxation introduced by Lynden-Bell \cite{lb}. The description of
coherent structures in Vlasov systems is complicated but it can be
explained in terms of ``classical'' principles without invoking a
generalized thermodynamics \cite{houches}.  Our discussion indicates
that there are two independent reasons why the quasi-equilibrium
states that form as a result of violent relaxation are
non-Boltzmannian. This is due, on the one hand, to the existence of
fine-grained constraints (the Casimirs) which depend on the initial
conditions and, on the other hand, to incomplete relaxation
(non-ergodicity, partial mixing). Even in case of ergodicity (complete
mixing), we can have a wide diversity of non-standard distributions
depending on the initial conditions.  They are given by
Eq. (\ref{E14}) according to the statistical theory of
Lynden-Bell. They are sorts of superstatistics. Moreover, if the
system does not mix efficiently, the Lynden-Bell prediction breaks
down and even more general distributions can be observed. They are
stable stationary solutions of the Vlasov equation on the
coarse-grained scale. The prediction of the metaequilibrium state in
case of incomplete relaxation is extremely complicated, if not
impossible. One possibility is to change the form of entropy. However,
the metaequilibrium state cannot apparently be described by a
universal functional such as the Tsallis functional, even if it is
extended to the form (\ref{se6}) so as to take into account the
specificities of the collisionless evolution (Casimir constraints). An
alternative approach is to keep the Lynden-Bell entropy but develop a
{\it dynamical} theory of violent relaxation as initiated in
\cite{csr,mn} to understand what prevents complete mixing. In that
case, we have to solve a dynamical equation with a non-constant
diffusion coefficient related to the fluctuations. The $H$-functions
can also be useful to construct stable models of galaxies (and 2D
vortices) in order to {\it reproduce} observed phenomena. In some
specific situations, some $H$-functions (belonging to the same ``class
of equivalence'') may be more appropriate than others to describe the
system, so that a phenomenological notion of ``effective generalized
thermodynamics'' (in $\overline{f}$-space or
$\overline{\omega}$-space) can be developed to deal with complex
systems in a simple and practical way
\cite{gfp}.  In that point of view, the relevant functional should be
found by trying and errors.

\appendix

\section{$H$-functions for the 2D Euler equation}
\label{app_h}

We briefly recall, and adapt to the case of the 2D Euler equation, the
notion of $H$-functions introduced by Tremaine {\it et al.}
\cite{thlb} for the Vlasov equation. These concepts 
 have not been introduced in 2D
turbulence. A $H$-function is a functional of the coarse-grained vorticity of the form
\begin{equation}
H=-\int C(\overline{\omega}) d^{2}{\bf r},
\label{ha1} 
\end{equation}
where $C$ is a convex function.  We assume that the initial condition at $t=0$ has been prepared without small-scale structure so that the fine-grained and coarse-grained vorticity fields are equal: $\overline{\omega}({\bf r},0)=\omega({\bf r},0)$. For $t>0$, the system will mix in a complicated manner and develop intermingled filaments so that these two fields will not be equal anymore. We have
\begin{eqnarray}
H(t)-H(0)=\int \lbrace C\lbrack \overline{\omega}({\bf r},0)\rbrack - C\lbrack \overline{\omega}({\bf r},t)\rbrack \rbrace d^{2}{\bf r}\nonumber\\
=\int \lbrace C\lbrack {\omega}({\bf r},0)\rbrack - C\lbrack \overline{\omega}({\bf r},t)\rbrack \rbrace d^{2}{\bf r}.
\label{ha2} 
\end{eqnarray}
The fine-grained vorticity is solution of the 2D Euler equation
\begin{eqnarray}
{\partial\omega\over\partial t}+{\bf u}\cdot \nabla\omega=0,
\label{ha3} 
\end{eqnarray}
where ${\bf u}({\bf r},t)$ is an incompressible velocity field. Thus
\begin{eqnarray}
{d\over dt}\int C(\omega)d^{2}{\bf r}=\int C'(\omega){\partial\omega\over\partial t}d^{2}{\bf r}=-\int C'(\omega){\bf u}\cdot\nabla\omega d^{2}{\bf r}\nonumber\\
=-\int {\bf u}\cdot\nabla C(\omega) d^{2}{\bf r}= -\int \nabla (C(\omega) {\bf u}) d^{2}{\bf r}=0.   
\label{ha4} 
\end{eqnarray}
This shows that the $H$-function $H\lbrack\omega\rbrack$ calculated
with the fine-grained vorticity is independent on time (it is a
particular Casimir) so Eq. (\ref{ha2}) becomes
\begin{eqnarray}
H(t)-H(0)=\int \lbrace C\lbrack {\omega}({\bf r},t)\rbrack - C\lbrack \overline{\omega}({\bf r},t)\rbrack \rbrace d^{2}{\bf r}.
\label{ha5} 
\end{eqnarray}
Now, a macrocell is divided into $\nu$ microcells of size $h=\Delta/\nu$. We call $\omega_{i}$ the value of the vorticity in a microcell. The contribution of a macrocell to $H(t)-H(0)$ is  
\begin{eqnarray}
\Delta\biggl\lbrace {1\over\nu}\sum_{i}C(\omega_{i})- C\biggl ({1\over\nu}\sum_{i} \omega_{i}\biggr )\biggr\rbrace
\label{ha6} 
\end{eqnarray}
which is positive since $C$ is convex. Therefore, the $H$-functions
calculated with the coarse-grained vorticity
$H\lbrack\overline{\omega}\rbrack$ increase in the sense that $H(t)\ge
H(0)$ for any $t\ge 0$. Note, however, that nothing is said concerning
the relative value of $H(t)$ and $H(t')$ for $t,t'>0$ so that the
increase is not necessarily monotonic. 

In 2D hydrodynamics, the viscosity has an effect similar
to coarse-graining. Indeed, considering the Navier-Stokes equation
\begin{eqnarray}
{\partial\omega\over\partial t}+{\bf u}\cdot \nabla\omega=\nu\Delta\omega,
\label{ha7} 
\end{eqnarray}
with $\nu>0$, we get 
\begin{eqnarray}
\dot H=-{d\over dt}\int C(\omega)d^{2}{\bf r}=-\int C'(\omega){\partial\omega\over\partial t}d^{2}{\bf r}=-\nu \int C'(\omega)\Delta\omega  d^{2}{\bf r}\nonumber\\
=\nu \int \nabla C'(\omega)\cdot \nabla\omega d^{2}{\bf r}= \nu \int C''(\omega)(\nabla\omega)^{2} d^{2}{\bf r}\ge 0.   
\label{ha8} 
\end{eqnarray}
In that case, the increase of $H$ is monotonic.


\begin{thebibliography}{7}
%
\addcontentsline{toc}{section}{References}


\bibitem{pais}  {\small A. Pais, {\it Subtle is the Lord}, Oxford University Press, New York (1982).}

\bibitem{cohen}  {\small E.G.D. Cohen, Physica A
{\bf 305}, 19 (2002).}

\bibitem{tsallis}  {\small C. Tsallis, J. Stat. Phys. {\bf 52}, 479
    (1988).}

\bibitem{collpap}  {\small Special issue of Physica A {\bf  340}, Issue 1-3, edited by G. Kaniadakis and M. Lissia (2004).}

\bibitem{bib}  {\small See http://tsallis.cat.cbpf.br/biblio.htm for a regularly updated bibliography on the subject.}

\bibitem{gfp}  {\small P.H. Chavanis, Phys. Rev. E {\bf 68}, 036108 (2003); P.H. Chavanis, Physica A {\bf 332}, 89 (2004); P.H. Chavanis, Banach Center Publ. {\bf 66}, 79 (2004); P.H. Chavanis, Physica A {\bf 340}, 57 (2004); P.H. Chavanis, P. Lauren\c cot, M. Lemou, Physica A \textbf{341}, 145 (2004).}

\bibitem{kaniadakis}  {\small G. Kaniadakis, Physica A {\bf 296}, 405 (2001).}

\bibitem{borland}  {\small L. Borland, Phys. Rev. E
{\bf 57}, 6634 (1998).}

\bibitem{frank}  {\small T.D. Frank, Physics Lett. A {\bf 290}, 93 (2001).}

\bibitem{bc}  {\small C. Beck \& E.G.D. Cohen, Physica A {\bf 322}, 267 (2003).}

\bibitem{souza}  {\small C. Tsallis \& A. Souza, Phys. Rev. E
{\bf 67}, 026106 (2003).}

\bibitem{lb}  {\small D. Lynden-Bell, Mon. Not. R. Astr. Soc.
{\bf 136}, 101 (1967).}

\bibitem{aa3}  {\small P.H. Chavanis, Astron. Astrophys.  {\bf 401}, 15 (2003).}

\bibitem{cs}  {\small P.H. Chavanis \& J. Sommeria, Mon. Not. R. Astr. Soc.
{\bf 296}, 569 (1998).}

\bibitem{miller}  {\small   J. Miller, Phys. Rev. Lett. {\bf 65}, 2137 (1990); R. Robert \& J. Sommeria, J. Fluid Mech. {\bf 229}, 291 (1991). }

\bibitem{csr}  {\small   P.H. Chavanis, J. Sommeria \& R. Robert, Astrophys. J. {\bf 471}, 385 (1996). }

\bibitem{houches}  {\small  P.H. Chavanis,  in {\it Dynamics and thermodynamics of systems with long range interactions}, edited by Dauxois, T, Ruffo, S., Arimondo, E. and  Wilkens, M. Lecture Notes in Physics, Springer (2002) [cond-mat/0212223].}

\bibitem{ellis}  {\small R. Ellis, K. Haven \& B. Turkington, Nonlinearity  {\bf 15}, 239 (2002).}

\bibitem{geo}  {\small P.H. Chavanis, Physica D {\bf 200}, 257 (2005). }

\bibitem{thlb}  {\small S. Tremaine, M. H\'enon \& D. Lynden-Bell, Mon. Not. R. Astron. Soc. {\bf 219}, 285
    (1986).}

\bibitem{cst}  {\small P.H. Chavanis \& C. Sire, [cond-mat/0409569] }

\bibitem{mn}  {\small P.H. Chavanis,  Mon. Not. R. Astr. Soc. {\bf 300}, 981 (1998).}

\bibitem{new}  {\small  P.H. Chavanis, [cond-mat/0409641]  }

\bibitem{meta}  {\small P.H. Chavanis, A\&A {\bf 432}, 117 (2005). }

\bibitem{bt}  {\small J. Binney and S. Tremaine,
{\it Galactic Dynamics} (Princeton Series in Astrophysics, 1987).}

\bibitem{lbw}  {\small  D. Lynden-Bell and R. Wood, Mon. Not. R. Astron. Soc. {\bf 138}, 495 (1968).}

\bibitem{kandrup}  {\small  H.E. Kandrup, Astrophys. J. {\bf 244}, 316 (1981).}

\bibitem{cb}  {\small  P.H. Chavanis and F. Bouchet,  Astron. Astrophys. {\bf 430}, 771 (2005) }

\bibitem{bellac}  {\small M. Le Bellac, {\it Quantum and Statistical Field Theory} (Clarendon Press, Oxford, 1991).}

\bibitem{cp}  {\small H.W. Capel and R.A. Pasmanter, Phys. Fluids   {\bf 12}, 
2514 (2000).}

\bibitem{almeida}  {\small M.P. Almeida, Physica A {\bf 300}, 424 (2001).}

\bibitem{beckn}  {\small C. Beck, Physica A {\bf 342}, 139 (2004).}

\bibitem{hm}  {\small J. Hjorth \& J. Madsen, Mon. Not. R. Astr. Soc. {\bf 265}, 237 (1993).}

\bibitem{brands}  {\small  H. Brands, P.H. Chavanis, J. Sommeria and R. Pasmanter, Phys. Fluids  {\bf 11}, 3465 (1999). }

\bibitem{onsager}  {\small L. Onsager, Nuovo Cimento Suppl. {\bf 6}, 279 (1949).}

\bibitem{jm}  {\small G. Joyce and  D. Montgomery, J. Plasma Phys. {\bf 10}, 107 (1973).}

\bibitem{dn}  {\small D. Dubin and T.M. O'Neil,  Phys. Rev. Lett. {\bf 60}, 1286 (1988); D. Dubin,  Phys. Plasmas {\bf 10}, 1338 (2003)}

\bibitem{pvkin}  {\small P.H. Chavanis, Phys. Rev. E {\bf 58}, R1199 (1998); P.H. Chavanis, Phys. Rev. E {\bf 64}, 026309 (2001).}

\bibitem{rsprl}  {\small R. Robert and J. Sommeria, Phys. Rev. Lett. {\bf 69}, 2776 (1992); R. Robert and C. Rosier, J. Stat. Phys. {\bf 86}, 481 (1997).}

\bibitem{hd}   {\small X.P. Huang and C.F. Driscoll,  Phys. Rev. Lett.  {\bf 72}, 2187 (1994).}

\bibitem{boghosian}   {\small B. Boghosian,  Phys. Rev. E  {\bf 53}, 4754 (1996).}

\bibitem{taruya}   {\small A. Taruya \& M. Sakagami,  Physica A {\bf 322}, 285 (2003)}

\bibitem{latora}  {\small  V. Latora, A. Rapisarda \& C. Tsallis, Physica A,  {\bf 305}, 129 (2002) }




\end{thebibliography}
\end{document}